\documentclass{article}
\usepackage{amsfonts}
\usepackage{makeidx}
\usepackage{latexsym,amsmath,amssymb,amscd}
\usepackage{makecell}
\usepackage{xcolor}
\usepackage{multirow}
\usepackage[all]{xy}
\usepackage{float}
\usepackage{longtable}

\allowdisplaybreaks
\newtheorem{theorem}{Theorem}

\newtheorem{proposition}[theorem]{Proposition}

\begin{document}

\title{Quadratic first integrals of time-dependent dynamical systems of the form $\ddot{q}^{a}= -\Gamma^{a}_{bc}\dot{q}^{b} \dot{q}^{c} -\omega(t)Q^{a}(q)$}
\author{Antonios Mitsopoulos$^{1,a)}$ and Michael Tsamparlis$^{1,b)}$ \\
{\ \ }\\
$^{1}${\textit{Faculty of Physics, Department of
Astronomy-Astrophysics-Mechanics,}}\\
{\ \textit{University of Athens, Panepistemiopolis, Athens 157 83, Greece}}
\vspace{12pt} 
\\
$^{a)}$Author to whom correspondence should be addressed: antmits@phys.uoa.gr 
\\
$^{b)}$Email: mtsampa@phys.uoa.gr }
\date{}
\maketitle

\begin{abstract}
We consider the time-dependent dynamical system
$\ddot{q}^{a}= -\Gamma_{bc}^{a}\dot{q}^{b}\dot{q}^{c}-\omega
(t)Q^{a}(q)$ where $\omega(t)$ is a non-zero arbitrary function and the connection coefficients $\Gamma^{a}_{bc}$ are computed from the kinetic metric (kinetic energy) of the system. In order to determine the quadratic first integrals (QFIs)
$I$ we assume that $I=K_{ab}\dot{q}^{a} \dot{q}^{b} +K_{a}\dot{q}^{a}+K$ where the unknown coefficients $K_{ab}, K_{a}, K$ are tensors depending on $t, q^{a}$ and impose the condition $\frac{dI}{dt}=0$. This condition leads to a system of partial differential equations (PDEs) involving the quantities $K_{ab}, K_{a}, K,$ $\omega(t)$ and $Q^{a}(q)$. From these PDEs, it follows that $K_{ab}$ is a Killing tensor (KT) of the kinetic metric. We use the KT $K_{ab}$ in two ways: a. We assume a general polynomial form in $t$ both for $K_{ab}$ and $K_{a}$; b. We express $K_{ab}$ in a basis of the KTs of order 2 of the kinetic metric assuming the coefficients to be functions of $t$. In both methods, this leads to a new system of PDEs whose solution requires that we specify either $\omega(t)$ or $Q^{a}(q)$. We consider first that $\omega(t)$ is a general polynomial in $t$ and find that in this case the dynamical system admits two independent QFIs which we collect in a Theorem. Next, we specify the quantities $Q^{a}(q)$ to be the generalized time-dependent Kepler potential $V=-\frac{\omega (t)}{r^{\nu}}$ and determine the functions $\omega(t)$ for which QFIs are admitted. We extend the discussion to the non-linear differential equation $\ddot{x}=-\omega(t)x^{\mu }+\phi (t)\dot{x}$ $(\mu \neq -1)$ and compute the relation between the coefficients $\omega(t), \phi(t)$ so that QFIs are admitted. We apply the results to determine the QFIs of the generalized Lane-Emden equation.
\end{abstract}

\section{Introduction}

\label{sec.intro}

The equations of motion of a dynamical system define in the configuration space a Riemannian structure with
the metric of the kinetic energy (kinetic metric). This metric is inherent in the
structure of the dynamical system; therefore, we expect that it will determine
the first integrals (FIs) of the system which are important in its evolution. On the other hand a metric is fixed  by
its symmetries, that is, the linear collineations: Killing vectors (KVs),
homothetic vectors (HVs), conformal Killing vectors (CKVs), affine collineations (ACs), projective collineations (PCs); and the quadratic
collineations: second order Killing tensors (KTs). The question then is how
the FIs of the dynamical system and the geometric symmetries of the kinetic
metric are related.

The standard way to determine the FIs of a differential equation is the use of Lie/Noether symmetries which applies to the point as well as the generalized Lie/Noether
symmetries. The relation of the Lie/Noether symmetries with the symmetries of the kinetic metric has been considered mostly in the case of point symmetries for autonomous conservative dynamical systems moving in a Riemannian space.
In particular, it has been shown (see e.g. \cite{Katzin 1974}, \cite{Tsamparlis 2011}, \cite{TsamparlisHV 2012}, \cite{AndrTsam 2012}) that the Lie point symmetries are generated by the special projective algebra of the kinetic metric whereas the Noether point symmetries are generated by the homothetic algebra of the kinetic metric, the latter being a subalgebra of the projective algebra. A recent clear statement of these results is discussed in \cite{Andr Tsamp 2015}.

In addition to the autonomous conservative systems this method has been applied to the time-dependent potentials $W(t,q)= \omega(t)V(q)$, that is, for equations of the form $\ddot{q}^{a}= -\Gamma_{bc}^{a} \dot{q}^{b} \dot{q}^{c} -\omega(t)V^{,a}(q)$ (see e.g. \cite{Katzin I1976}, \cite{Katzin II1977}, \cite{Katzin 1977}, \cite{Ray 1979B}, \cite{Prince 1980}, \cite{Ray 1980}, \cite{LeoTsampAndro 2017}). In
this case it has been shown that the Lie point symmetries, the Noether point symmetries and the
associated FIs are computed in terms of the collineations of the kinetic
metric plus a set of constraint conditions involving the time-dependent
potential and the collineation vectors. These time-dependent potentials are
important because (among others) they contain the time-dependent oscillator (see e.g. \cite{Katzin 1977}, \cite{Prince 1980}, \cite{Lewis 1968}, \cite{Gunther 1977}, \cite{Ray 1979A}) and the time-dependent Kepler potential (see e.g. \cite{LeoTsampAndro 2017}, \cite{Prince 1981}, \cite{Katzin 1982}, \cite{Leach 1985}).
A further development in the same line  is the extension of this method to time-dependent potentials $W(t,q)$ with linear damping terms \cite{LeoTsampAndro 2017}. It has been shown that under a suitable time
transformation the damping term can be removed and the problem reduces to a
time-dependent potential of the form $W(t,q)=\bar{\omega}(t)V(q)$ but
with different $\bar{\omega}(t)$. Finally the Lie/Noether method has been applied to
the study of partial differential equations (PDEs) \cite{AndrTsam 2012}, \cite{RoseKatizn 1994}, \cite{Bozhkov 2010}, \cite{Tsamp 2015}.

Besides the aforementioned Lie/Noether method there is a different method which computes the FIs in terms of the collineations of the kinetic metric without using Lie symmetries. This method we shall apply in this paper. It has as follows.

One assumes the generic quadratic first integral (QFI) to be of the form\footnote{The linear FIs (LFIs) are also included for $K_{ab}=0$.}
\begin{equation}
I=K_{ab}\dot{q}^{a}\dot{q}^{b}+K_{a}\dot{q}^{a}+K \label{FL.5}
\end{equation}%
where the coefficients $K_{ab}, K_{a}, K$ are tensors depending on the coordinates $t, q^{a}$ and imposes the condition $\frac{dI}{dt}=0.$ Using
again the equations of motion to replace the quantities $\ddot{q}^{a}$
whenever they appear, this condition leads to a system of PDEs involving the unknown quantities $K_{ab}, K_{a}, K$ and the
dynamical elements, i.e. the potential and the generalized forces of the
system. The solution of this system of PDEs provides the QFIs (\ref{FL.5}). For future reference we shall call this method the {\emph{direct method}}.

The system of PDEs consists of two parts: a. The geometric part which is independent of the dynamical quantities; and b. the dynamical part which contains the scalar $K$ and the
dynamical quantities. The main conclusion of the geometric part is that  the tensor $K_{ab}$ is a KT of the kinetic metric whereas the vector $K_{a}$ is related to the linear collineations of that metric. The dynamical part involves the scalar $K$ which is determined by a set of
constraint conditions which involve $K_{ab}, K_{a}, K$, the potential and the
generalized forces. Once $K$ is computed one gets the corresponding QFI $I$.

The direct method can always be related to the Noether
symmetries. Indeed assuming that the system has a regular Lagrangian (which is always the case since we assume that there exists the kinetic energy) it can be shown by using the inverse Noether theorem (see \cite{Djukic 1975} and section II in \cite{Mitsopoulos 2020}) that to each QFI $I$ one determines an associated gauged generalized Noether symmetry with generator $\eta_{a}= -2K_{ab}\dot{q}^{b} -K_{a}$ and Noether function $f= -K_{ab}\dot{q}^{a}\dot{q}^{b} +K$ whose Noether integral is the considered QFI. Therefore we conclude that all QFIs of the form (\ref{FL.5}) are
Noetherian, provided the Lagrangian is regular, that is, the dynamical equations can be solved in terms of $\ddot{q}^{a}$.

Moreover, this method has  been employed in the literature (see \cite{Katzin 1982}, \cite{Katzin 1973}, \cite{Katzin 1981}, \cite{Katzin 1983}) both for autonomous and time-dependent dynamical systems. A recent account of this method in the case of autonomous conservative systems together with relevant references can be found in \cite{Tsamparlis 2020}. This approach being geometric is powerful and convenient
because with minimal calculations it allows the computation of the FIs by
using known results from differential geometry.

The purpose of the present work is to apply the direct method to compute the QFIs of time-dependent equations of the form  $\ddot{q}^{a} =-\Gamma_{bc}^{a} \dot{q}^{b}\dot{q}^{c} -\omega(t)Q^{a}(q)$. Because many well-known dynamical systems fall in this category we intend to recover in a direct single approach all the known results derived from the Lie/Noether symmetry method, which are scattered in a large number of papers.

As explained above, the solution of the system requires that the tensor $K_{ab}$ is a KT of the kinetic metric.
In general, the computation of the KTs of a metric is a major task. However for spaces of constant curvature this problem has been solved (see \cite{Thompson 1984}, \cite{Thompson 1986}, \cite{Horwood 2008}). Therefore, in this paper, we restrict our discussion to Euclidean spaces only. Since the KT $K_{ab}$ is a function of $t,q^a$ we suggest two procedures of work: a. The polynomial method; b. the basis method.

In the polynomial method, one assumes a general polynomial form in the variable $t$ both for the KT $K_{ab}$ and the vector $K_{a}$ and replaces in the equations of the relevant system. In the basis method, one computes first a basis of the KTs of order 2 of the kinetic
metric and then expresses in this basis the KT $K_{ab}$ with the coefficients to be functions of $t$. The vector $K_{a}$ and the FIs follow from the solution of the system. Both methods are suitable for autonomous dynamical systems but for time-dependent systems it appears that the basis method is preferable.

Concerning the quantities $\omega(t)$ and $Q^{a}(q)$, again, there are two ways to proceed. \newline
a) Consider a general form for the function $\omega (t)$ and let the quantities $Q^{a}$ unspecified. In this case the quantities $Q^{a}$ act as constraints.\newline
b) Specify the quantities $Q^{a}$ and determine for which functions $\omega(t)$ the resulting dynamical system admits QFIs.

In the following we shall consider both the polynomial method and the basis method, starting from the former. As a first application, we assume the KT $K_{ab}=N(t)\gamma_{ab}$ where $N(t)$ is an arbitrary function and show that we recover all the point Noether integrals found in \cite{LeoTsampAndro 2017}. As a second application, we assume that $\omega(t) =b_{0}+b_{1}t+...+b_{\ell }t^{\ell }$ with $b_{\ell }\neq 0$ and $\ell \geq 1$ whereas the quantities $Q^{a}$ are unspecified. We find that in this case the system admits two families of independent QFIs as stated in Theorem \ref{thm.polynomial.omega}.

Subsequently, we consider the basis method. This is carried out in two steps. In the first step, we assume that we know a basis $\{C_{(N)ab}(q)\}$ of the space of KTs of the kinetic metric and require that $K_{ab}$ has the form $K_{ab}(t,q) =\sum_{N=1}^{m}\alpha _{N}(t)C_{(N)ab}(q)$. In the second ste,p we specify the generalized forces to be conservative with the time-dependent Newtonian generalized Kepler potential $V=-\frac{\omega (t)}{r^{\nu}}$ where $\nu$ is a non-zero real constant and $r=\sqrt{x^{2}+y^{2}+z^{2}}$. This potential for $\nu=-2,1$ includes respectively the three-dimensional (3d) time-dependent oscillator and the time-dependent Kepler potential. For other values of $\nu$ it reduces to other important dynamical systems, for example, for $\nu=2$ one obtains the Newton-Cotes potential (see e.g. \cite{Ibragimov 1998}). We determine the QFIs of the time-dependent generalized Kepler potential and recover in a systematic way the known results concerning the QFIs of the 3d time-dependent oscillator, the time-dependent Kepler potential  and the Newton-Cotes potential. For easier reference we collect all the results in Table \ref{T2} of section \ref{Table 2}.

Using the well-known result that by a reparameterization the linear damping term $\phi(t)\dot{q}^{a}$ of a dynamical equation is absorbed to a time-dependent force of the form $\omega(t)Q^{a}(q)$, we also study the non-linear differential equation $\ddot{x}=-\omega(t)x^{\mu }+\phi (t)\dot{x}$ $(\mu \neq -1)$ and compute the relation between the coefficients $\omega(t), \phi(t)$ for which QFIs are admitted. It is found that a family of `frequencies' $\bar{\omega}(s)$ is admitted which for $\mu =0, 1, 2$ is parameterized with functions whereas for $\mu \neq -1,0,1,2$ is parameterized with constants.  As a further application, we study the integrability of the well-known generalized Lane-Emden equation.

The structure of the paper is as follows. In section \ref{sec.conditions} we determine the system of PDEs resulting form the condition $dI/dt=0$.  In section \ref{sec.point.integrals}, we assume that the KT is proportional to the kinetic metric and derive the point Noether FIs of the time-dependent dynamical system (\ref{eq.red1}). In section \ref{sec.direct.solution}, we consider the polynomial method and define the general forms of the KT $K_{ab}$ and the vector $K_{a}$ which lead to a new form of the system of PDEs. In section \ref{sec.polynomial.omega}, we assume that $\omega(t)$ is a general polynomial of $t$ and we find that the resulting time-dependent system admits two independent QFIs as stated in Theorem \ref{thm.polynomial.omega}. In section \ref{sec.In}, we discuss some special cases of the QFI $I_{n}$ of Theorem \ref{thm.polynomial.omega}. In section \ref{sec.general.approach}, we consider the basis method. In section \ref{sec.KTE3}, we find a basis for the KTs in $E^{3}$ in order to apply the basis method to 3d Newtonian systems. In sections \ref{sec.TimeGKepler} - \ref{sec.discussion.GK}, we study the time-dependent generalized Kepler potential and find for which functions $\omega(t)$ admits QFIs. Particularly, in section \ref{sec.discussion.GK}, we study a special class of time-dependent oscillators with frequency $\omega_{3O}(t)$ as given in equation (\ref{eq.disosc3}). We collect our results for the several values of $\nu$ in Table \ref{T2} of section \ref{Table 2}. In section \ref{sec.applications.GK}, we use the independent LFIs $I_{41i}, I_{42i}$ given in equations (\ref{eq.disosc4b}), (\ref{eq.disosc4c}) to integrate the equations of the time-dependent oscillators defined in section \ref{sec.discussion.GK}; and the FIs $L_{i}$, $E_{2}$, $A_{i}$ determined in subsection \ref{sec.omega.2} to integrate the time-dependent Kepler potential with $\omega(t)= \frac{k}{b_{0}+b_{1}t}$ where $kb_{1}\neq0$. In section \ref{sec.nonlin}, we consider the second order non-linear time-dependent differential equation (\ref{eq.nonl1}) and show that it is integrable with an associated QFI given in equation (\ref{eq.nonl6.2}) iff the functions $\omega(t), \phi(t)$ are related as shown in equation (\ref{eq.nonl6.1}). For the special values $\mu=0,1,2$ we find also that there exist additional relations between $\omega(t), \phi(t)$ for which the resulting differential equation admits a QFI. For $\mu=1$ equation (\ref{eq.nonl1}) admits the general solution (\ref{eq.nonl4.2.7}) provided that condition (\ref{eq.nonl4.2.6}) is satisfied. We apply these results in subsection \ref{sec.emden} and we study the properties of the well-known generalized Lane-Emden equation. Finally, in section \ref{conclusions}, we draw our conclusions and, in the appendix, we give the proof of Theorem \ref{thm.polynomial.omega}.

\section{The system of equations}

\label{sec.conditions}

We consider the dynamical system
\begin{equation}
\ddot{q}^{a} = - \Gamma^{a}_{bc} \dot{q}^{b} \dot{q}^{c} - \omega(t)Q^{a}(q) \label{eq.red1}
\end{equation}
where $\Gamma^{a}_{bc}$ are the Riemannian connection coefficients determined by the kinetic metric $\gamma_{ab}$ (kinetic energy) of the system and $-\omega(t)Q^{a}(q)$ are the time-dependent generalized forces. Einstein summation convention is assumed and the metric $\gamma_{ab}$ is used for lowering and raising the indices.

We next consider a function $I(t,q^{a},\dot{q}^{a})$ of the form
\begin{equation}
I=K_{ab}(t,q)\dot{q}^{a}\dot{q}^{b}+K_{a}(t,q)\dot{q}^{a}+K(t,q)
\label{FI.5}
\end{equation}%
where $K_{ab}$ is a symmetric tensor, $K_{a}$ is a vector and $K$ is an invariant.

We demand $I$ be a FI of (\ref{eq.red1}) by imposing the
condition
\begin{equation}
\frac{dI}{dt}=0.  \label{DS1.10a}
\end{equation}%
Using the dynamical equations (\ref{eq.red1}) to replace $\ddot{q}^{a}$ whenever it appears we find\footnote{%
Round brackets indicate symmetrization of the enclosed indices. A comma
indicates partial derivative and a semicolon Riemannian covariant derivative.%
} the system of equations
\begin{eqnarray}
K_{(ab;c)} &=&0  \label{eq.TKN1} \\
K_{ab,t}+K_{(a;b)} &=&0  \label{eq.TKN2} \\
-2\omega K_{ab}Q^{b}+K_{a,t}+K_{,a} &=&0  \label{eq.TKN3} \\
K_{,t}-\omega K_{a}Q^{a} &=&0  \label{eq.TKN4} \\
K_{a,tt}+\omega \left( K_{b}Q^{b}\right) _{,a}-2\omega
_{,t}K_{ab}Q^{b}-2\omega K_{ab,t}Q^{b} &=&0  \label{eq.TKN5} \\
K_{[a;b],t}-2\omega \left( K_{[a|c|}Q^{c}\right)_{;b]} &=&0
\label{eq.TKN6}
\end{eqnarray}
where the last two equations\footnote{
These equations come from the conditions $K_{,[at]}=0$ and $K_{,[ab]}=0$ respectively.} (\ref{eq.TKN5}), (\ref{eq.TKN6}) express the integrability conditions for the scalar $K$.

Equation (\ref{eq.TKN1}) implies that $K_{ab}$ is a KT of order 2 (possibly zero) of the kinetic metric $\gamma_{ab}$.

The solution of the system requires the function $\omega(t)$ and the quantities $Q^{a}(q)$ both being quantities which are characteristic of the given dynamical system. There are two ways to proceed. \newline
a) Consider a general form for the function $\omega(t)$ and let the quantities $Q^{a}(q)$
unspecified. In this case the
quantities $Q^{a}(q)$ act as constraints.\newline
b) Specify the quantities $Q^{a}(q)$ and determine for which functions $\omega(t)$ the resulting dynamical system admits FIs.

However, before continuing with this kind of considerations, we first proceed with the simple geometric choice $K_{ab}=N(t)\gamma_{ab}$ where $N(t)$ is an arbitrary smooth function. By specifying the KT $K_{ab}$ as above both the function $\omega(t)$ and the quantities $Q^{a}(q)$ stay unspecified and can act as constraints.

\section{The point Noether FIs of the time-dependent dynamical system (\ref{eq.red1})}

\label{sec.point.integrals}

We consider the simplest choice
\begin{equation}
K_{ab}=N(t)\gamma_{ab} \label{eq.KTtr}
\end{equation}
where $N(t)$ is an arbitrary smooth function. This choice is purely geometric; therefore,  the function $\omega(t)$ and the quantities $Q^{a}(q)$ are unspecified and act as constraints, whereas the vector $K_{a}$ is identified with one collineation of the kinetic metric. With this $K_{ab}$, the system of equations (\ref{eq.TKN1}) - (\ref{eq.TKN6}) become (eq. (\ref{eq.TKN1}) vanishes trivially)
\begin{eqnarray}
N_{,t}\gamma_{ab} +K_{(a;b)} &=&0  \label{eq.Tpoint.16a} \\
-2\omega N Q_{a}+K_{a,t}+K_{,a} &=&0  \label{eq.Tpoint.16b} \\
K_{,t}-\omega K_{a}Q^{a} &=&0  \label{eq.Tpoint.16c} \\
K_{a,tt} +\omega\left(K_{b}Q^{b}\right)_{,a} -2\omega_{,t}NQ_{a} -2\omega N_{,t}Q_{a} &=& 0 \label{eq.Tpoint.16d} \\
K_{[a;b],t} -2 \omega NQ_{[a;b]} &=&0.  \label{eq.Tpoint.16e}
\end{eqnarray}

We consider the following cases.

\subsection{Case $K_{a}=K_{a}(q)$ is the HV of $\gamma_{ab}$ with homothety factor $\psi$}

\label{sec.p1}

In this case $K_{a,t}=0$ and $K_{(a;b)}= \psi\gamma_{ab}$ where $\psi$ is an arbitrary constant.

Equation (\ref{eq.Tpoint.16a}) gives
\[
N_{,t}=-\psi \implies N=-\psi t + c
\]
where $c$ is an arbitrary constant.

Equation (\ref{eq.Tpoint.16e}) implies that (take $\omega \neq 0$)
\[
Q_{[a;b]}=0 \implies Q_{a}=V_{,a}
\]
where $V=V(q)$ is an arbitrary potential.

Replacing in (\ref{eq.Tpoint.16b}) we find that
\[
K_{,a} =2\omega (-\psi t +c)V_{,a} \implies K= 2\omega (-\psi t +c)V + M(t)
\]
where $M(t)$ is an arbitrary function.

Substituting the function $K(t,q)$ in (\ref{eq.Tpoint.16c}) we get
\begin{equation}
\omega K_{a}V^{,a} - 2\omega_{,t} (-\psi t +c)V + 2\omega \psi V -M_{,t} =0. \label{eq.Tpoint.17}
\end{equation}
The remaining condition (\ref{eq.Tpoint.16d}) is just the partial derivative of (\ref{eq.Tpoint.17}), and hence is satisfied trivially.

Moreover, since $\omega \neq 0$, equation (\ref{eq.Tpoint.17}) can be written in the form
\begin{equation}
K_{a}V^{,a} - 2(\ln\omega)_{,t} (-\psi t +c)V + 2\psi V -\frac{M_{,t}}{\omega} =0 \label{eq.Tpoint.18}
\end{equation}
which implies that
\begin{eqnarray}
2(\ln\omega)_{,t} (-\psi t +c) &=& c_{1} \label{eq.Tpoint.18a} \\
M_{,t} &=& c_{2}\omega \label{eq.Tpoint.18b}
\end{eqnarray}
where $c_{1}, c_{2}$ are arbitrary constants.

Therefore equation (\ref{eq.Tpoint.18}) becomes
\begin{equation}
K_{a}V^{,a} +(2\psi-c_{1}) V -c_{2} =0. \label{eq.Tpoint.18c}
\end{equation}

The QFI is
\begin{equation}
I_{1} = (-\psi t+c)\gamma_{ab}\dot{q}^{a}\dot{q}^{b} + K_{a}(q)\dot{q}^{a} + 2\omega (-\psi t +c)V + M(t) \label{eq.Tpoint.18d}
\end{equation}
where $Q_{a}=V_{,a}$ and the quantities $\omega(t), M(t), V(q), K_{a}(q)$ satisfy the conditions (\ref{eq.Tpoint.18a}) - (\ref{eq.Tpoint.18c}).

\subsection{Case $K_{a}= -M(t)S_{,a}(q)$ where $S_{,a}$ is the gradient HV of $\gamma_{ab}$}

\label{sec.p2}

In this case $S_{;ab}=\psi \gamma_{ab}$ and $M(t)\neq0$ is an arbitrary function.

Equation (\ref{eq.Tpoint.16a}) implies $N_{,t}= \psi M$.

From equation (\ref{eq.Tpoint.16e}) we find that there exists a potential function $V(q)$ such that $Q_{a}=V_{,a}$.

Replacing the above results in (\ref{eq.Tpoint.16b}) we obtain
\[
K_{,a}= 2\omega NV_{,a} +M_{,t}S_{,a} \implies K= 2\omega NV + M_{,t}S + C(t)
\]
where $C(t)$ is an arbitrary function.

Substituting in (\ref{eq.Tpoint.16c}) we get (take $\omega M\neq 0$)
\[
\omega M S_{,a}V^{,a} +2\omega_{,t} NV + 2\omega \psi MV + M_{,tt}S + C_{,t} =0 \implies S_{,a}V^{,a} + 2\psi V + \frac{2(\ln\omega)_{,t} N}{M} V + \frac{M_{,tt}}{\omega M}S + \frac{C_{,t}}{\omega M} =0
\]
which implies that
\begin{eqnarray}
\frac{2(\ln\omega)_{,t} N}{M} &=& d_{1} \label{eq.Tpoint.19a} \\
\frac{M_{,tt}}{\omega M} &=& m \label{eq.Tpoint.19b} \\
\frac{C_{,t}}{\omega M} &=& k \label{eq.Tpoint.19c} \\
S_{,a}V^{,a} + (2\psi + d_{1})V + mS + k &=& 0 \label{eq.Tpoint.19d}
\end{eqnarray}
where $d_{1}, m, k$ are arbitrary constants. The remaining condition (\ref{eq.Tpoint.16d}) is satisfied identically.

The QFI is
\begin{equation}
I_{2} = N\gamma_{ab} \dot{q}^{a}\dot{q}^{b} - MS_{,a}\dot{q}^{a} +2\omega NV + M_{,t}S + C(t) \label{eq.Tpoint.19e}
\end{equation}
where $Q_{a}=V_{,a}$, $N_{,t}= \psi M$ and the conditions (\ref{eq.Tpoint.19a}) - (\ref{eq.Tpoint.19d}) must be satisfied.

\subsection{Case $Q_{a}=V_{,a}$ and $K_{a}= -M(t)V_{,a}(q)$ where $V_{,a}$ is the gradient HV of $\gamma_{ab}$}

\label{sec.p3}

Equation (\ref{eq.Tpoint.16a}) implies $N_{,t}= \psi M$ where $\psi$ is the homothety factor of $V_{,a}$.

From equation (\ref{eq.Tpoint.16b}) we obtain
\[
K_{,a}= 2\omega NV_{,a} +M_{,t}V_{,a} \implies K= 2\omega NV + M_{,t}V + C(t)
\]
where $C(t)$ is an arbitrary function.

Substituting in (\ref{eq.Tpoint.16c}) we get (take $\omega M\neq 0$)
\[
\omega M V_{,a}V^{,a} +2\omega_{,t} NV + 2\omega \psi MV + M_{,tt}V + C_{,t} =0 \implies V_{,a}V^{,a} + 2\psi V + \frac{2(\ln\omega)_{,t} N}{M} V + \frac{M_{,tt}}{\omega M}V + \frac{C_{,t}}{\omega M} =0
\]
which implies that
\begin{eqnarray}
\frac{M_{,tt}}{\omega M} + \frac{2(\ln\omega)_{,t} N}{M} &=& d_{2} \label{eq.Tpoint.20a} \\
\frac{C_{,t}}{\omega M} &=& k \label{eq.Tpoint.20b} \\
V_{,a}V^{,a} + (2\psi + d_{2})V + k &=& 0 \label{eq.Tpoint.20c}
\end{eqnarray}
where $d_{2}, k$ are arbitrary constants. The remaining conditions are satisfied identically.

The QFI is
\begin{equation}
I_{3} = N\gamma_{ab} \dot{q}^{a}\dot{q}^{b} - MV_{,a}\dot{q}^{a} + \left( 2\omega N + M_{,t} \right)V + C \label{eq.Tpoint.20d}
\end{equation}
where $Q_{a}=V_{,a}$, $N_{,t}= \psi M$ and the conditions (\ref{eq.Tpoint.20a}) - (\ref{eq.Tpoint.20c}) must be satisfied.
\bigskip

The above results reproduce Theorem 2 of \cite{LeoTsampAndro 2017} which states that the point Noether symmetries
of the time-dependent potentials of the form $\omega(t)V(q)$ are generated by the homothetic algebra of the kinetic metric (provided the Lagrangian is regular).

It is interesting to observe that the QFIs (\ref{eq.Tpoint.18d}), (\ref{eq.Tpoint.19e}), (\ref{eq.Tpoint.20d}) produced by point Noether symmetries can be also produced by generalized (gauged) Noether symmetries using the inverse Noether theorem. This proves that a Noether FI is not associated with a unique Noether symmetry!

\section{The polynomial method for computing the QFIs}

\label{sec.direct.solution}

In the polynomial approach one assumes a polynomial form in $t$ of the KT $K_{ab}(t,q)$ and the vector $K_{a}(t,q)$ and solves the resulting system for given $\omega(t), Q^{a}(q)$. One
application of this method can be found in \cite{Tsamparlis 2020} where a
general theorem is given which allows the finding of the QFIs of an
autonomous conservative dynamical system. In the present work we generalize the considerations
made in \cite{Tsamparlis 2020} and assume that the quantity $K_{ab}(t,q)$ has the form
\begin{equation}
K_{ab}(t,q)=C_{(0)ab}(q) + \sum_{N=1}^{n}C_{(N)ab}(q)\frac{t^{N}}{N} \label{eq.aspm1}
\end{equation}%
where $C_{(N)ab}$, $N=0,1,...,n$, is a sequence of arbitrary KTs of order 2 of the kinetic metric $\gamma_{ab}$.

This choice of $K_{ab}$ and equation (\ref{eq.TKN2}) indicate that we set
\begin{equation}
K_{a}(t,q)= \sum^{m}_{M=0} L_{(M)a}(q) t^{M} \label{eq.aspm2}
\end{equation}%
where $L_{(M)a}(q)$, $M=0,1,...,m$, are arbitrary vectors.

We note that both powers $n$, $m$ in the above polynomial expressions may be infinite.

Substituting (\ref{eq.aspm1}), (\ref{eq.aspm2}) in the system of equations (\ref{eq.TKN1}) - (\ref{eq.TKN6}) (equation (\ref{eq.TKN1}) is identically zero since $C_{(N)ab}$ are KTs) we obtain the system of equations
\begin{eqnarray}
0&=& C_{(1)ab} + C_{(2)ab}t + ... + C_{(n)ab} t^{n-1} + L_{(0)(a;b)} + L_{(1)(a;b)}t +... + L_{(m)(a;b)}t^{m} \label{eq.red2a} \\
0 &=& -2\omega C_{(0)ab}Q^{b} -2\omega C_{(1)ab}Q^{b} t - ... - 2\omega C_{(n)ab} Q^{b} \frac{t^{n}}{n} + L_{(1)a} + 2L_{(2)a}t + ... + mL_{(m)a}t^{m-1} + K_{,a} \label{eq.red2b} \\
0&=& K_{,t}- \omega L_{(0)a}Q^{a} - \omega L_{(1)a}Q^{a}t - ... - \omega L_{(m)a}Q^{a}t^{m} \label{eq.red2c} \\
0 &=& \left(-2C_{(0)ab}Q^{b} -2C_{(1)ab}Q^{b} t - ... - 2C_{(n)ab} Q^{b} \frac{t^{n}}{n}\right) \omega_{,t} -2\omega C_{(1)ab}Q^{b} -2\omega C_{(2)ab}Q^{b}t -... - 2 \omega C_{(n)ab}Q^{b}t^{n-1} + \notag \\
&& + 2L_{(2)a}+6L_{(3)a}t+...+m(m-1)L_{(m)a}t^{m-2}+ \omega\left( L_{(0)b}Q^{b}\right) _{,a}+ \omega \left( L_{(1)b}Q^{b}\right)_{,a}t +...+ \omega \left( L_{(m)b}Q^{b}\right)_{,a}t^{m} \label{eq.red2d} \\
0 &=&2\omega\left( C_{(0)[a\left\vert c\right\vert }Q^{c}\right) _{;b]}+ 2\omega \left(
C_{(1)[a\left\vert c\right\vert }Q^{c}\right) _{;b]}t + ...+ 2\omega \left(C_{(n)[a\left\vert c\right\vert }Q^{c}\right) _{;b]}\frac{t^{n}}{n} - L_{(1)\left[a;b\right] } - \notag \\
&&- 2L_{(2)\left[ a;b\right] }t-...-mL_{(m)\left[ a;b\right] }t^{m-1}. \label{eq.red2e}
\end{eqnarray}

In this system of PDEs the pairs $\omega(t), Q^{a}(q)$ are not specified. As we explained in the introduction we shall fix a general form of $\omega$ and find the admitted QFIs in terms of the (unspecified) $Q^a$. In the following section we choose $\omega(t)$ to be a general polynomial in $t$, however any other choice is possible.

\section{The case $\mathbf{\omega(t)=b_{0} +b_{1}t + ... + b_{\ell}t^{\ell}}$ with $\mathbf{b_{\ell}\neq0}$, $\mathbf{\ell\geq1}$}

\label{sec.polynomial.omega}

We assume that
\begin{equation}
\omega(t)= b_{0}+b_{1}t+ ... + b_{\ell}t^{\ell}, \enskip b_{\ell}\neq0, \enskip \ell \geq 1 \label{pol}
\end{equation}
where $\ell$ is the degree of the polynomial. Substituting the function (\ref{pol}) in the system of equations (\ref{eq.red2a}) - (\ref{eq.red2e}) we find\footnote{The proof of Theorem \ref{thm.polynomial.omega} is in the appendix.} that there are two independent QFIs as given in Theorem \ref{thm.polynomial.omega}.

\begin{theorem}
\label{thm.polynomial.omega}
The independent QFIs of the time-dependent dynamical system (\ref{eq.red1}) where $\omega(t)= b_{0} +b_{1}t + ... +b_{\ell}t^{\ell }$ with $b_{\ell }\neq 0$ and $\ell \geq 1$ are the following:
\bigskip

\textbf{Integral 1.}

\begin{eqnarray*}
I_{n}&=&  \left( C_{(0)ab} +\sum^{n}_{k=1} \frac{t^{k}}{k} C_{(k)ab} \right) \dot{q}^{a} \dot{q}^{b} + \sum^{n}_{k=0} t^{k} L_{(k)a} \dot{q}^{a} +\sum^{n}_{k=0} \sum^{\ell}_{r=0} \left( L_{(k)a}Q^{a} b_{r}\frac{t^{k+r+1}}{k+r+1} \right) +G(q)
\end{eqnarray*}
where $n=0, 1, 2, ...$, $C_{(0)ab}$ is a KT, the KTs $C_{(N)ab} = -L_{(N-1)(a;b)}$ for $N=1,...,n$, $L_{(n)a}$ is a KV, $G(q)$ is an arbitrary function defined by the condition
\begin{equation}
G_{,a}= 2b_{0}C_{(0)ab}Q^{b} -L_{(1)a} \label{thm0}
\end{equation}
$s$ is an arbitrary constant defined by the condition
\begin{equation}
L_{(n)a}Q^{a}=s \label{thm1}
\end{equation}
and the following conditions are satisfied
\begin{equation}
\sum_{s=0}^{\ell-1}\left[  -\frac{2(r+s) b_{(r+s\leq\ell)}}{n-s} C_{(n-s\geq0)ab}Q^{b} -2b_{(r+s\leq\ell)} C_{(n-s>0)ab}Q^{b} + b_{(r+s\leq\ell)} \left( L_{(n-s-1\geq0)b}Q^{b}\right)_{,a} \right]=0, \enskip r=1,2,...,\ell \label{thm2}
\end{equation}
\begin{equation}
-\sum_{s=1}^{\ell}\left[ \frac{2sb_{s}}{n-s} C_{(n-s\geq0)ab}Q^{b} \right] + \sum_{s=0}^{\ell} \left[ -2b_{s} C_{(n-s>0)ab}Q^{b} + b_{s} \left(L_{(n-s-1\geq0)b}Q^{b} \right)_{,a} \right]=0 \label{thm3}
\end{equation}
\begin{equation}
k(k-1)L_{(k)a} - \sum_{s=1}^{\ell} \left[ \frac{2sb_{s}}{k-s-1} C_{(k-s-1\geq0)ab}Q^{b} \right] + \sum_{s=0}^{\ell} \left[ -2b_{s}C_{(k-s-1>0)ab}Q^{b} +b_{s}\left( L_{(k-s-2\geq0)b}Q^{b} \right)_{,a} \right] =0 \label{thm4}
\end{equation}
with $k=2,3,...n$.

\textbf{Integral 2.}

\begin{equation*}
I_{e} =I_{e}(\ell=1) = - e^{\lambda t} L_{(a;b)} \dot{q}^{a}\dot{q}^{b} + \lambda e^{\lambda t} L_{a}\dot{q}^{a} + \left( b_{0} - \frac{b_{1}}{\lambda} \right) e^{\lambda t}L_{a}Q^{a} + b_{1}te^{\lambda t} L_{a}Q^{a}
\end{equation*}%
where $L_{(a;b)}$ is a KT, $\left( L_b Q^{b}\right)_{,a}= \frac{\lambda^{3}}{b_{1}}L_{a}$ and $\lambda^3 L_a = -2b_{1} L_{(a;b)} Q^{b}$.

We note that the FI $I_{e}$ exists only when $\omega(t)=b_{0}+b_{1}t$, that is, for $\ell=1$.
\end{theorem}

\section{Special cases of the QFI $I_{n}$}

\label{sec.In}

The parameter $n$ in the case Integral 1 of Theorem \ref{thm.polynomial.omega} runs over all positive integers, i.e. $n=0,1,2,...$. This results in a sequence of QFIs $I_{0}, I_{1}, I_{2}, ...$, one QFI $I_{n}$ for each value $n$. A significant characteristic of this sequence is that $I_{k} < I_{k+1}$, that is, each QFI $I_{k}$ where $k=0,1,2,...$ can be derived from the next QFI $I_{k+1}$ as a subcase.

In the following we consider some special cases of the QFI $I_{n}$ for small values of $n$.

\subsection{The QFI $I_{0}$}

For $n=0$ we have
\begin{equation*}
I_{0}= C_{(0)ab} \dot{q}^{a} \dot{q}^{b} +L_{(0)a}\dot{q}^{a} + b_{\ell}s \frac{t^{\ell+1}}{\ell+1} + ... + b_{1}s \frac{t^{2}}{2} + b_{0}st
\end{equation*}
where $C_{(0)ab}$ is a KT, $L_{(0)a}$ is a KV, $L_{(0)a}Q^{a}=s$ and $C_{(0)ab}Q^{b}=0$.

This QFI consists of the independent FIs
\[
I_{0a}= C_{(0)ab} \dot{q}^{a} \dot{q}^{b}, \enskip I_{0b}= L_{(0)a}\dot{q}^{a} + b_{\ell}s \frac{t^{\ell+1}}{\ell+1} + ... + b_{1}s \frac{t^{2}}{2} + b_{0}st.
\]

\subsection{The QFI $I_{1}$}

For $n=1$ the conditions (\ref{thm1}) - (\ref{thm4}) become
\begin{eqnarray}
L_{(1)a}Q^{a}&=& s \label{eq.polc0.2} \\
\left( L_{(0)b} Q^{b} \right)_{,a} &=& -2(\ell+1)L_{(0)(a;b)} Q^{b} \label{eq.polc0.3} \\
kb_{k} C_{(0)ab}Q^{b} &=& - (\ell-k+1)b_{k-1}L_{(0)(a;b)}Q^{b}, \enskip k=1,...,\ell. \label{eq.polc0.4}
\end{eqnarray}

Since $b_{\ell}\neq 0$ the last condition for $k=\ell$ gives
\[
C_{(0)ab}Q^{b} = - \frac{b_{\ell-1}}{\ell b_{\ell}} L_{(0)(a;b)} Q^{b}
\]
and the remaining equations become
\[
\left[(\ell-k+1)b_{k-1} -\frac{kb_{k}b_{\ell-1}}{\ell b_{\ell}} \right] L_{(0)(a;b)} Q^{b} =  0, \enskip k=1,...,\ell-1.
\]
The last set of equations exist only for $\ell\geq2$. From these equations, using mathematical induction, we prove after successive substitutions that
\[
\left( b_{0} - \frac{b^{\ell}_{\ell-1}}{\ell^{\ell} b^{\ell-1}_{\ell}} \right) L_{(0)(a;b)} Q^{b}=0.
\]

The QFI is ($I_{0}$ is a subcase of $I_{1}$)
\begin{eqnarray*}
I_{1} &=& \left( -tL_{(0)(a;b)} + C_{(0)ab} \right) \dot{q}^{a} \dot{q}^{b} + tL_{(1)a}\dot{q}^{a} +L_{(0)a}\dot{q}^{a} +sb_{\ell}\frac{t^{\ell+2}}{\ell+2} + \left( sb_{\ell-1} + b_{\ell}L_{(0)a}Q^{a} \right) \frac{t^{\ell+1}}{\ell+1} + ... + \\
&& + \left( sb_{0} + b_{1}L_{(0)a}Q^{a} \right) \frac{t^{2}}{2} + b_{0}L_{(0)a}Q^{a}t + G(q)
\end{eqnarray*}
where $C_{(0)ab}$, $L_{(0)(a;b)}$ are KTs, $L_{(1)a}$ is a KV, $L_{(1)a}Q^{a}=s$, $\left( L_{(0)b} Q^{b} \right)_{,a} = -2(\ell+1)L_{(0)(a;b)}Q^{b}$, $C_{(0)ab}Q^{b} = - \frac{b_{\ell-1}}{\ell b_{\ell}} L_{(0)(a;b)} Q^{b}$, $\left[(\ell-k+1)b_{k-1} -\frac{kb_{k}b_{\ell-1}}{\ell b_{\ell}} \right] L_{(0)(a;b)} Q^{b} =0$ where  $k=1,...,\ell-1$ and $G_{,a}= 2b_{0}C_{(0)ab}Q^{b} - L_{(1)a}$.
\bigskip

For some values of the degree $\ell$ of the polynomial $\omega(t)$ we have:

1) For $\ell=1$.

We have $\omega=b_{0}+b_{1}t$ and the QFI is
\begin{equation*}
I_{1}= \left( -tL_{(0)(a;b)} + C_{(0)ab} \right) \dot{q}^{a} \dot{q}^{b} + tL_{(1)a}\dot{q}^{a} +L_{(0)a}\dot{q}^{a} +sb_{1}\frac{t^{3}}{3} + \left( sb_{0} + b_{1}L_{(0)a}Q^{a} \right) \frac{t^{2}}{2} + b_{0}L_{(0)a}Q^{a}t + G(q)
\end{equation*}
where $C_{(0)ab}$, $L_{(0)(a;b)}$ are KTs, $L_{(1)a}$ is a KV, $L_{(1)a}Q^{a}=s$, $\left( L_{(0)b} Q^{b} \right)_{,a} = -4L_{(0)(a;b)}Q^{b}$, $C_{(0)ab}Q^{b} = - \frac{b_{0}}{b_{1}} L_{(0)(a;b)} Q^{b}$ and $G_{,a}= 2b_{0}C_{(0)ab}Q^{b} - L_{(1)a}$.

2) For $\ell=2$.

We have $\omega=b_{0}+b_{1}t+b_{2}t^{2}$ and the QFI is
\begin{eqnarray*}
I_{1} &=& \left( -tL_{(0)(a;b)} + C_{(0)ab} \right) \dot{q}^{a} \dot{q}^{b} + tL_{(1)a}\dot{q}^{a} +L_{(0)a}\dot{q}^{a} +sb_{2}\frac{t^{4}}{4} + \left( sb_{1} + b_{2}L_{(0)a}Q^{a} \right) \frac{t^{3}}{3} \\
&& + \left( sb_{0} + b_{1}L_{(0)a}Q^{a} \right) \frac{t^{2}}{2} + b_{0}L_{(0)a}Q^{a}t + G(q)
\end{eqnarray*}
where $C_{(0)ab}$, $L_{(0)(a;b)}$ are KTs, $L_{(1)a}$ is a KV, $L_{(1)a}Q^{a}=s$, $\left( L_{(0)b} Q^{b} \right)_{,a} = -6L_{(0)(a;b)}Q^{b}$, $C_{(0)ab}Q^{b} = - \frac{b_{1}}{2b_{2}} L_{(0)(a;b)} Q^{b}$, $\left( b_{0} -\frac{b_{1}^{2}}{4b_{2}} \right) L_{(0)(a;b)} Q^{b} =0$ and $G_{,a}= 2b_{0}C_{(0)ab}Q^{b} - L_{(1)a}$.

3) For $\ell=3$.

We have $\omega=b_{0} + b_{1}t + b_{2}t^{2}+ b_{3}t^{3}$ and the QFI is
\begin{eqnarray*}
I_{1} &=& \left( -tL_{(0)(a;b)} + C_{(0)ab} \right) \dot{q}^{a} \dot{q}^{b} + tL_{(1)a}\dot{q}^{a} +L_{(0)a}\dot{q}^{a} +sb_{3}\frac{t^{5}}{5} + \left( sb_{2} + b_{3}L_{(0)a}Q^{a} \right) \frac{t^{4}}{4} + \left( sb_{1} + b_{2}L_{(0)a}Q^{a} \right) \frac{t^{3}}{3} + \\
&& + \left( sb_{0} + b_{1}L_{(0)a}Q^{a} \right) \frac{t^{2}}{2} + b_{0}L_{(0)a}Q^{a}t + G(q)
\end{eqnarray*}
where $C_{(0)ab}$, $L_{(0)(a;b)}$ are KTs, $L_{(1)a}$ is a KV, $L_{(1)a}Q^{a}=s$, $\left( L_{(0)b} Q^{b} \right)_{,a} = -8L_{(0)(a;b)}Q^{b}$, $C_{(0)ab}Q^{b} = - \frac{b_{2}}{3b_{3}} L_{(0)(a;b)} Q^{b}$, $\left( b_{0} - \frac{b_{1}b_{2}}{9b_{3}} \right) L_{(0)(a;b)} Q^{b} =0$, $\left( b_{1} - \frac{b_{2}^{2}}{3b_{3}} \right) L_{(0)(a;b)} Q^{b} =0$ and $G_{,a}= 2b_{0}C_{(0)ab}Q^{b} - L_{(1)a}$.

\section{The basis method for computing QFIs}

\label{sec.general.approach}

As it has been explained in the introduction, in the basis method instead of
considering the KT \thinspace $K_{ab}$ to be given as a polynomial in $t$
with coefficients arbitrary KTs (see equation (\ref{eq.aspm1}) ) one defines the KT $K_{ab}(t,q)$ by the requirement
\begin{equation}
K_{ab}(t,q)=\sum_{N=1}^{m}\alpha_{N}(t)C_{(N)ab}(q)  \label{eq.secapr.1}
\end{equation}%
where $\alpha_{N}(t)$ are arbitrary smooth functions and the $m$ linearly independent KTs $C_{(N)ab}(q)$ constitute a basis of the space of KTs
of the kinetic metric $\gamma_{ab}(q)$. In this case, one does not assume a form for the vector
$K_{a}(t,q)$ which is determined from the resulting system of equations (\ref{eq.TKN1}) - (\ref{eq.TKN6}).

The basis method has been used previously by Katzin and
Levine in \cite{Katzin 1982} in order to determine the QFIs for the time-dependent Kepler potential. As we shall apply the basis method to 3d Newtonian systems we need a basis of KTs (and other collineations) of the Euclidean space $E^{3}$.

\section{The geometric quantities of $E^{3}$}

\label{sec.KTE3}

In $E^{3}$ the general KT of order 2 has independent components%
\begin{eqnarray}
C_{11} &=&\frac{a_{6}}{2}y^{2}+\frac{a_{1}}{2}%
z^{2}+a_{4}yz+a_{5}y+a_{2}z+a_{3}  \notag \\
C_{12} &=&\frac{a_{10}}{2}z^{2}-\frac{a_{6}}{2}xy-\frac{a_{4}}{2}xz-\frac{%
a_{14}}{2}yz-\frac{a_{5}}{2}x-\frac{a_{15}}{2}y+a_{16}z+a_{17}  \notag \\
C_{13} &=&\frac{a_{14}}{2}y^{2}-\frac{a_{4}}{2}xy-\frac{a_{1}}{2}xz-\frac{%
a_{10}}{2}yz-\frac{a_{2}}{2}x+a_{18}y-\frac{a_{11}}{2}z+a_{19}  \label{FL.E3}
\\
C_{22} &=&\frac{a_{6}}{2}x^{2}+\frac{a_{7}}{2}%
z^{2}+a_{14}xz+a_{15}x+a_{12}z+a_{13}  \notag \\
C_{23} &=&\frac{a_{4}}{2}x^{2}-\frac{a_{14}}{2}xy-\frac{a_{10}}{2}xz-\frac{%
a_{7}}{2}yz-(a_{16}+a_{18})x-\frac{a_{12}}{2}y-\frac{a_{8}}{2}z+a_{20}
\notag \\
C_{33} &=&\frac{a_{1}}{2}x^{2}+\frac{a_{7}}{2}%
y^{2}+a_{10}xy+a_{11}x+a_{8}y+a_{9}  \notag
\end{eqnarray}%
where $a_{I}$ with $I=1,2,...,20$ are arbitrary real constants.

The vector $L^{a}$ generating the KT $C_{ab}=L_{(a;b)}$ is
\begin{equation}
L_{a}=\left(
\begin{array}{c}
-a_{15}y^{2}-a_{11}z^{2}+a_{5}xy+a_{2}xz+2(a_{16}+a_{18})yz+a_{3}x
+2a_{4}y+2a_{1}z+a_{6} \\
-a_{5}x^{2}-a_{8}z^{2}+a_{15}xy-2a_{18}xz+a_{12}yz+
2(a_{17}-a_{4})x+a_{13}y+2a_{7}z+a_{14} \\
-a_{2}x^{2}-a_{12}y^{2}-2a_{16}xy+a_{11}xz+a_{8}yz+2(a_{19}-
a_{1})x+2(a_{20}-a_{7})y+a_{9}z+a_{10}%
\end{array}%
\right)  \label{eq.Kep.5}
\end{equation}
and the generated KT is
\begin{equation}
C_{ab}=\left(
\begin{array}{ccc}
a_{5}y+a_{2}z+a_{3} & -\frac{a_{5}}{2}x-\frac{a_{15}}{2}y+a_{16}z+a_{17} & -%
\frac{a_{2}}{2}x+a_{18}y-\frac{a_{11}}{2}z+a_{19} \\
-\frac{a_{5}}{2}x-\frac{a_{15}}{2}y+a_{16}z+a_{17} & a_{15}x+a_{12}z+a_{13}
& -(a_{16}+a_{18})x-\frac{a_{12}}{2}y-\frac{a_{8}}{2}z+a_{20} \\
-\frac{a_{2}}{2}x+a_{18}y-\frac{a_{11}}{2}z+a_{19} & -(a_{16}+a_{18})x-\frac{%
a_{12}}{2}y-\frac{a_{8}}{2}z+a_{20} & a_{11}x+a_{8}y+a_{9}%
\end{array}%
\right)  \label{eq.Kep.8}
\end{equation}%
which is a subcase of the general KT (\ref{FL.E3}) for $%
a_{1}=a_{4}=a_{6}=a_{7}=a_{10}=a_{14}=0$.

We note that the covariant expression of the most general KT $M_{ij}$
of order 2 of $E^{3}$ is (see \cite{Crampin 1984}, \cite{Chanu 2006})
\begin{equation}
M_{ij}=(\varepsilon _{ikm}\varepsilon _{jln}+\varepsilon
_{jkm}\varepsilon _{iln})A^{mn}q^{k}q^{l}+(B_{(i}^{l}\varepsilon
_{j)kl}+\lambda _{(i}\delta _{j)k}-\delta _{ij}\lambda _{k})q^{k}+D_{ij}
\label{CRA.46}
\end{equation}%
where $A^{mn}, B_{i}^{l}, D_{ij}$ are constant tensors all being symmetric and $B_{i}^{l}$ also being traceless; $\lambda^{k}$ is a constant vector; and $\varepsilon_{ijk}$ is the 3d Levi-Civita symbol. This
result is obtained from the solution of the Killing tensor equation in Euclidean space.

Observe that $A^{mn}$, $D_{ij}$ have each 6 independent components; $%
B_{i}^{l}$ has 5 independent components; and $\lambda ^{k}$ has 3
independent components. Therefore $M_{ij}$ depends on $6+6+5+3=20$
arbitrary real constants, a result which is in accordance with the one given above in equation (\ref{FL.E3}).

\section{The time-dependent Newtonian generalized Kepler potential}

\label{sec.TimeGKepler}

The time-dependent Newtonian generalized Kepler potential is $V=-\frac{\omega(t)}{r^{\nu}}$ where $\nu$ is a non-zero real constant and $r=(x^{2}+y^{2}+z^{2})^{\frac{1}{2}}$. This potential contains (among others) the 3d time-dependent oscillator \cite{Katzin 1977}, \cite{Prince 1980}, \cite{Lewis 1968}, \cite{Gunther 1977}, \cite{Ray 1979A} for $\nu =-2$, the time-dependent Kepler potential \cite{LeoTsampAndro 2017}, \cite{Prince 1981}, \cite{Katzin 1982}, \cite{Leach 1985} for $\nu =1$ and the Newton-Cotes potential for $\nu=2$ \cite{Ibragimov 1998}.  The integrability of these systems has been studied in numerous works
over the years using various methods, mainly the Noether symmetries. Our purpose is to recover the results of these works - and also new ones - using the basis
method.

The Lagrangian of the system is
\begin{equation}
L=\frac{1}{2}(\dot{x}^{2}+\dot{y}^{2}+\dot{z}^{2})+\frac{\omega(t) }{r^{\nu}}
\label{eq.TGKep.1}
\end{equation}%
and the corresponding Euler-Lagrange equations are
\begin{equation}
\ddot{x}=-\frac{\nu \omega(t)}{r^{\nu+2}}x, \enskip \ddot{y}=-\frac{\nu \omega(t)}{r^{\nu +2}}y, \enskip\ddot{z}=-\frac{\nu \omega(t)}{r^{\nu +2}}z.
\label{eq.TGKep.1a}
\end{equation}
For this system the $Q^{a}= \frac{\nu q^{a}}{r^{\nu+2}}$ where $q^{a}= (x,y,z)$ whereas the $\omega(t)$ is unspecified. We shall determine those $\omega(t)$ for which the resulting FIs are not combinations of the angular
momentum.

The LFIs and the QFIs of the autonomous generalized Kepler potential, that is, $\omega(t)=k=const$, have been determined in \cite{Tsamparlis 2020} using the direct method and are listed in Table \ref{T1}.

\begin{longtable}{|l|l|}
\hline
$V=-\frac{k}{r^{\nu}}$ & LFIs and QFIs \\ \hline
$\forall$ $\nu$ & $L_{1} = y\dot{z} - z\dot{y}$, $L_{2}= z\dot{x} - x\dot{z}$, $L_{3}= x\dot{y} - y\dot{x}$, $H_{\nu}= \frac{1}{2}(\dot{x}^{2} + \dot{y}^{2} + \dot{z}^{2}) - \frac{k}{r^{\nu}}$ \\
$\nu=-2$ & $B_{ij} = \dot{q}_{i} \dot{q}_{j} - 2k q_{i}q_{j}$ \\
$\nu=-2$, $k>0$ & $I_{3a\pm}= e^{\pm \sqrt{2k} t}(\dot{q}_{a} \mp \sqrt{2k} q_{a})$ \\
$\nu=-2$, $k<0$ & $I_{3a\pm}= e^{\pm i \sqrt{-2k} t}(\dot{q}_{a} \mp i \sqrt{-2k} q_{a})$ \\
$\nu=1$ & $R_{i}= (\dot{q}^{j} \dot{q}_{j}) q_{i} - (\dot{q}^{j}q_{j})\dot{q}_{i}- \frac{k}{r}q_{i}$ \\
$\nu=2$ & $I_{1}= - H_{2}t^{2} + t(\dot{q}^{i}q_{i}) - \frac{r^{2}}{2}$, $I_{2}= - H_{2}t + \frac{1}{2} (\dot{q}^{i}q_{i})$
 \\ \hline
\caption{\label{T1} The LFIs/QFIs of the autonomous generalized Kepler potential for $\omega(t)=k=const$.}
\end{longtable}

In Table \ref{T1} $H_{\nu}$ is the Hamiltonian of the system, $L_{i}$ are the components of the angular momentum, $R_{i}$ are the components of the Runge-Lenz vector and $B_{ij}$ are the components of the Jauch-Hill-Fradkin tensor.
\bigskip

Using $Q^{a}=\frac{\nu q^{a}}{r^{\nu+2}}$ conditions (\ref{eq.TKN1}) - (\ref{eq.TKN6}) become (see \cite{Katzin 1982})
\begin{eqnarray}
K_{(ab;c)} &=&0  \label{eq.TKNq1} \\
K_{(a;b)} + K_{ab,t} &=& 0 \label{eq.TKNq2} \\
K_{,a} -\frac{2\nu\omega}{r^{\nu+2}}K_{ab}q^{b} +K_{a,t} &=&0 \label{eq.TKNq3} \\
K_{,t} -\frac{\nu \omega}{r^{\nu+2}}K_{a}q^{a} &=&0 \label{eq.TKNq4} \\
K_{a,tt} + \nu\omega\left( \frac{K_{b}q^{b}}{r^{\nu+2}} \right)_{,a} -\frac{2\nu\omega_{,t}}{r^{\nu+2}}K_{ab}q^{b} -\frac{2\nu\omega}{r^{\nu+2}}K_{ab,t}q^{b} &=& 0 \label{eq.TKNq5} \\
K_{[a;b],t} -2 \nu\omega\left( \frac{K_{[a|c|}q^{c}}{r^{\nu+2}} \right)_{;b]} &=&0.  \label{eq.TKNq6}
\end{eqnarray}

From the Lagrangian (\ref{eq.TGKep.1}) we infer that the kinetic metric is $\delta_{ij}=diag(1,1,1)$.

According to the basis approach, the KT $K_{ab}(t,q)$ of (\ref{eq.TKNq1}) is the KT given by (\ref{FL.E3}) but the 20 arbitrary constants $a_{I}$ are assumed to be time-dependent functions $a_{I}(t)$.

Condition (\ref{eq.TKNq2}) gives
\[
K_{a,b} + K_{b,a} = -2K_{ab,t} \implies
\]
\begin{eqnarray}
K_{1,1} &=& -K_{11,t} \label{eq.TKNq7a} \\
K_{2,2} &=& -K_{22,t} \label{eq.TKNq7b} \\
K_{3,3} &=& -K_{33,t} \label{eq.TKNq7c} \\
K_{1,2} + K_{2,1} &=& -2K_{12,t} \label{eq.TKNq7d} \\
K_{1,3} + K_{3,1} &=& -2K_{13,t} \label{eq.TKNq7e} \\
K_{2,3} + K_{3,2} &=& -2K_{23,t}. \label{eq.TKNq7f}
\end{eqnarray}

From the first three conditions (\ref{eq.TKNq7a}) - (\ref{eq.TKNq7c}) we find
\begin{eqnarray*}
K_{1} &=& -\frac{\dot{a}_{6}}{2}xy^{2} -\frac{\dot{a}_{1}}{2}
xz^{2} -\dot{a}_{4}xyz -\dot{a}_{5}xy -\dot{a}_{2}xz -\dot{a}_{3}x +A(y,z,t) \\
K_{2} &=& -\frac{\dot{a}_{6}}{2}yx^{2} -\frac{\dot{a}_{7}}{2} yz^{2} -\dot{a}_{14}xyz -\dot{a}_{15}xy -\dot{a}_{12}yz -\dot{a}_{13}y + B(x,z,t) \\
K_{3} &=& -\frac{\dot{a}_{1}}{2}zx^{2} -\frac{\dot{a}_{7}}{2}zy^{2} -\dot{a}_{10}xyz -\dot{a}_{11}xz -\dot{a}_{8}yz -\dot{a}_{9}z + C(x,y,t)
\end{eqnarray*}
where $A, B, C$ are arbitrary functions.

Substituting these results in (\ref{eq.TKNq7d}) - (\ref{eq.TKNq7f}) we obtain
\begin{eqnarray}
0&=& \dot{a}_{10}z^{2} -3\dot{a}_{6}xy -2\dot{a}_{4}xz -2\dot{a}_{14}yz -2\dot{a}_{5}x  -2\dot{a}_{15}y +2\dot{a}_{16}z +2\dot{a}_{17} + A_{,2} + B_{,1} \label{eq.TKNq8a} \\
0&=& \dot{a}_{14}y^{2} -2\dot{a}_{4}xy -3\dot{a}_{1}xz -2\dot{a}_{10}yz -2\dot{a}_{2}x +2\dot{a}_{18}y -2\dot{a}_{11}z +2\dot{a}_{19} + A_{,3} + C_{,1} \label{eq.TKNq8b}
\\
0&=& \dot{a}_{4}x^{2} -2\dot{a}_{14}xy -2\dot{a}_{10}xz -3\dot{a}_{7}yz -2(\dot{a}_{16}+\dot{a}_{18})x -2\dot{a}_{12}y -2\dot{a}_{8}z +2\dot{a}_{20} + B_{,3} + C_{,2}. \label{eq.TKNq8c}
\end{eqnarray}

By taking the second partial derivatives of (\ref{eq.TKNq8a}) with respect to (wrt) $x, y$, of (\ref{eq.TKNq8b}) wrt $x, z$ and of (\ref{eq.TKNq8c}) wrt $y,z$ we find that
\[
a_{1}=c_{1}, \enskip a_{6}=c_{2}, \enskip a_{7}=c_{3}
\]
are arbitrary constants.

Then equations (\ref{eq.TKNq8a}) - (\ref{eq.TKNq8c}) become
\begin{eqnarray}
0&=& \dot{a}_{10}z^{2} -2\dot{a}_{4}xz -2\dot{a}_{14}yz -2\dot{a}_{5}x  -2\dot{a}_{15}y +2\dot{a}_{16}z +2\dot{a}_{17} + A_{,2} + B_{,1} \label{eq.TKNq9a} \\
0&=& \dot{a}_{14}y^{2} -2\dot{a}_{4}xy -2\dot{a}_{10}yz -2\dot{a}_{2}x +2\dot{a}_{18}y -2\dot{a}_{11}z +2\dot{a}_{19} + A_{,3} + C_{,1} \label{eq.TKNq9b} \\
0&=& \dot{a}_{4}x^{2} -2\dot{a}_{14}xy -2\dot{a}_{10}xz -2(\dot{a}_{16}+\dot{a}_{18})x -2\dot{a}_{12}y -2\dot{a}_{8}z +2\dot{a}_{20} + B_{,3} + C_{,2}. \label{eq.TKNq9c}
\end{eqnarray}

By suitable differentiations of the above equations we obtain
\begin{eqnarray*}
A_{,22} &=& 2\dot{a}_{14}z +2\dot{a}_{15} \\
A_{,33} &=& 2\dot{a}_{10}y + 2\dot{a}_{11} \\
B_{,11} &=& 2\dot{a}_{4}z + 2\dot{a}_{5} \\
B_{,33}&=& 2\dot{a}_{10}x +2\dot{a}_{8} \\
C_{,11} &=& 2\dot{a}_{4}y + 2\dot{a}_{2} \\
C_{,22} &=& 2\dot{a}_{14}x +2\dot{a}_{12}.
\end{eqnarray*}
Then
\begin{eqnarray*}
A&=& \dot{a}_{14}zy^{2} +\dot{a}_{10}yz^{2} +\dot{a}_{15}y^{2} + \dot{a}_{11}z^{2} + \sigma_{1}(t)yz + \sigma_{2}(t)y + \sigma_{3}(t)z + \sigma_{4}(t) \\
B&=& \dot{a}_{4}zx^{2} + \dot{a}_{10}xz^{2} + \dot{a}_{5}x^{2} + \dot{a}_{8}z^{2} + \tau_{1}(t)xz + \tau_{2}(t)x + \tau_{3}(t)z + \tau_{4}(t) \\
C&=& \dot{a}_{4}yx^{2} + \dot{a}_{14}xy^{2} + \dot{a}_{2}x^{2} + \dot{a}_{12}y^{2} + \eta_{1}(t)xy + \eta_{2}(t)x + \eta_{3}(t)y + \eta_{4}(t)
\end{eqnarray*}
where $\sigma_{k}(t), \tau_{k}(t), \eta_{k}(t)$ for $k=1,2,3,4$ are arbitrary functions.

Substituting in (\ref{eq.TKNq9a}) - (\ref{eq.TKNq9c}) we find
\begin{eqnarray*}
(\ref{eq.TKNq9a}) &\implies& a_{10}=c_{4}, \enskip \sigma_{1}= -\tau_{1} -2\dot{a}_{16}, \enskip \sigma_{2}= - \tau_{2} -2\dot{a}_{17} \\
(\ref{eq.TKNq9b}) &\implies& a_{14}=c_{5}, \enskip \eta_{1}= -\sigma_{1} -2\dot{a}_{18}, \enskip \eta_{2}= - \sigma_{3} -2\dot{a}_{19} \\
(\ref{eq.TKNq9c}) &\implies& a_{4}=c_{6}, \enskip \tau_{1}= -\eta_{1} +2(\dot{a}_{16} + \dot{a}_{18}), \enskip \tau_{3}= -\eta_{3} -2\dot{a}_{20}
\end{eqnarray*}
from which we have finally
\[
a_{10}=c_{4}, \enskip a_{14}=c_{5}, \enskip a_{4}=c_{6}, \enskip \tau_{1}=2\dot{a}_{18}, \enskip \eta_{1}=2\dot{a}_{16}, \enskip \sigma_{1}= -2(\dot{a}_{16} + \dot{a}_{18}),
\]
\[
\tau_{2}= - \sigma_{2} -2\dot{a}_{17}, \enskip \eta_{2}= - \sigma_{3} -2\dot{a}_{19}, \enskip \eta_{3}= -\tau_{3} -2\dot{a}_{20}
\]
where $c_{4}, c_{5}, c_{6}$ are arbitrary constants.

Therefore the KT $K_{ab}$ is
\begin{eqnarray}
K_{11} &=&\frac{c_{2}}{2}y^{2}+\frac{c_{1}}{2}%
z^{2}+c_{6}yz+a_{5}y+a_{2}z+a_{3} \notag \\
K_{12} &=&\frac{c_{4}}{2}z^{2}-\frac{c_{2}}{2}xy-\frac{c_{6}}{2}xz-\frac{%
c_{5}}{2}yz-\frac{a_{5}}{2}x-\frac{a_{15}}{2}y+a_{16}z+a_{17}  \notag \\
K_{13} &=&\frac{c_{5}}{2}y^{2}-\frac{c_{6}}{2}xy-\frac{c_{1}}{2}xz-\frac{%
c_{4}}{2}yz-\frac{a_{2}}{2}x+a_{18}y-\frac{a_{11}}{2}z+a_{19} \label{eq.KT}
\\
K_{22} &=&\frac{c_{2}}{2}x^{2}+\frac{c_{3}}{2}%
z^{2} +c_{5}xz+a_{15}x+a_{12}z+a_{13} \notag \\
K_{23} &=&\frac{c_{6}}{2}x^{2}-\frac{c_{5}}{2}xy -\frac{c_{4}}{2}xz-\frac{c_{3}}{2}yz -(a_{16}+a_{18})x-\frac{a_{12}}{2}y -\frac{a_{8}}{2}z+a_{20} \notag \\
K_{33} &=&\frac{c_{1}}{2}x^{2}+\frac{c_{3}}{2}%
y^{2}+c_{4}xy+a_{11}x+a_{8}y+a_{9} \notag
\end{eqnarray}
and the vector $K_{a}$ is
\begin{eqnarray}
K_{1} &=& \dot{a}_{15}y^{2} + \dot{a}_{11}z^{2} -\dot{a}_{5}xy -\dot{a}_{2}xz -2(\dot{a}_{16}+\dot{a}_{18})yz -\dot{a}_{3}x + \sigma_{2}y + \sigma_{3}z + \sigma_{4} \notag \\
K_{2} &=& \dot{a}_{5}x^{2} + \dot{a}_{8}z^{2} -\dot{a}_{15}xy + 2\dot{a}_{18}xz -\dot{a}_{12}yz -(\sigma_{2}+2\dot{a}_{17})x -\dot{a}_{13}y + \tau_{3}z + \tau_{4} \label{eq.K} \\
K_{3} &=& \dot{a}_{2}x^{2} + \dot{a}_{12}y^{2} +2\dot{a}_{16}xy -\dot{a}_{11}xz -\dot{a}_{8}yz -(\sigma_{3}+2\dot{a}_{19})x -(\tau_{3}+2\dot{a}_{20})y -\dot{a}_{9}z + \eta_{4}. \notag
\end{eqnarray}

Replacing the above results in the constraint (\ref{eq.TKNq6}) we find the following set of equations:
\begin{equation}
a_{2}=a_{12}, \enskip a_{5}=a_{8}, \enskip a_{11}=a_{15}, \enskip a_{16}=a_{18}=0 \label{eq.conTK1a}
\end{equation}
\begin{equation}
(\nu-1)a_{2}=0, \enskip (\nu-1)a_{5}=0, \enskip (\nu-1)a_{11}=0 \label{eq.conTK1b}
\end{equation}
\begin{equation}
(\nu+2)a_{17}=0, \enskip (\nu+2)a_{19}=0, \enskip (\nu+2)a_{20}=0, \enskip (\nu+2)(a_{3}-a_{9})=0, \enskip (\nu+2)(a_{3}-a_{13})=0 \label{eq.conTK1c}
\end{equation}
\begin{equation}
\ddot{a}_{2}=\ddot{a}_{5}=\ddot{a}_{11}=0, \enskip \dot{\sigma}_{2}= -\ddot{a}_{17}, \enskip \dot{\sigma}_{3}= -\ddot{a}_{19}, \enskip \dot{\tau}_{3}= -\ddot{a}_{20}. \label{eq.conTK1d}
\end{equation}

We consider three cases depending on the value of $\nu$:

- $\forall \nu$. The general case.

- $\nu =1$. Time-dependent Kepler potential.

- $\nu =-2$. Time-dependent 3d oscillator.

The Newton-Cotes potential ($\nu=2$) is contained as a subcase of the general case.

\section{The general case}

\label{3.2}

This case holds for any value of $\nu$ and conditions (\ref{eq.conTK1a}) - (\ref{eq.conTK1d}) give
\[
a_{2}=a_{5}=a_{8}=a_{11}=a_{12}=a_{15}=a_{16}=a_{17} =a_{18}=a_{19}=a_{20}=0,
\]
\[
a_{3}=a_{9}=a_{13}, \enskip \sigma_{2}=c_{7}, \enskip \sigma_{3}= c_{8}, \enskip \tau_{3}= c_{9}
\]
where $c_{7}, c_{8}, c_{9}$ are arbitrary constants.

Substituting in the constraint (\ref{eq.TKNq5}) we find that
\begin{equation}
\dddot{a}_{3}=0, \enskip (\nu-2)\omega\dot{a}_{3} - 2\dot{\omega}a_{3}=0 \label{eq.gen1.1}
\end{equation}
\[
\ddot{\sigma}_{4}= \ddot{\tau}_{4}= \ddot{\eta}_{4}=0, \enskip \omega\sigma_{4}=\omega\tau_{4}=\omega\eta_{4}=0 \implies \sigma_{4}=\tau_{4}=\eta_{4}=0.
\]

Therefore the KT $K_{ab}$ becomes
\begin{equation}
K_{ab}=
\left(
  \begin{array}{ccc}
    \frac{c_{2}}{2}y^{2}+\frac{c_{1}}{2}z^{2}+c_{6}yz+a_{3} & \frac{c_{4}}{2}z^{2}-\frac{c_{2}}{2}xy-\frac{c_{6}}{2}xz-\frac{%
c_{5}}{2}yz & \frac{c_{5}}{2}y^{2}-\frac{c_{6}}{2}xy-\frac{c_{1}}{2}xz-\frac{%
c_{4}}{2}yz \\
    \frac{c_{4}}{2}z^{2}-\frac{c_{2}}{2}xy-\frac{c_{6}}{2}xz-\frac{%
c_{5}}{2}yz & \frac{c_{2}}{2}x^{2}+\frac{c_{3}}{2}%
z^{2} +c_{5}xz +a_{3} & \frac{c_{6}}{2}x^{2}-\frac{c_{5}}{2}xy -\frac{c_{4}}{2}xz-\frac{c_{3}}{2}yz \\
    \frac{c_{5}}{2}y^{2}-\frac{c_{6}}{2}xy-\frac{c_{1}}{2}xz-\frac{%
c_{4}}{2}yz & \frac{c_{6}}{2}x^{2}-\frac{c_{5}}{2}xy -\frac{c_{4}}{2}xz-\frac{c_{3}}{2}yz & \frac{c_{1}}{2}x^{2} +\frac{c_{3}}{2} y^{2}+c_{4}xy +a_{3} \\
  \end{array}
\right) \label{eq.gencas.1}
\end{equation}
and the vector
\begin{equation}
K_{a}=
\left(
  \begin{array}{c}
    -\dot{a}_{3}x + c_{7}y + c_{8}z \\
    -c_{7}x -\dot{a}_{3}y + c_{9}z \\
    -c_{8}x -c_{9}y -\dot{a}_{3}z \\
  \end{array}
\right). \label{eq.gencas.2}
\end{equation}

Since the ten parameters $a_{3}(t)$ and $c_{A}$ where $A=1, 2, ..., 9$ are independent (i.e. they generate different FIs) we consider the following two cases.

\subsection{$a_{3}(t)=0$}

\label{sec.gen1}

In this case the conditions (\ref{eq.gen1.1}) are satisfied identically leaving the function $\omega(t)$ free to be any function.

Therefore the KT (\ref{eq.gencas.1}) becomes
\begin{equation*}
K_{ab}=
\left(
\begin{array}{ccc}
\frac{c_{2}}{2}y^{2}+\frac{c_{1}}{2}z^{2}+c_{6}yz & \frac{c_{4}}{2}z^{2}-\frac{c_{2}}{2}xy-\frac{c_{6}}{2}xz-\frac{%
c_{5}}{2}yz & \frac{c_{5}}{2}y^{2}-\frac{c_{6}}{2}xy-\frac{c_{1}}{2}xz-\frac{%
c_{4}}{2}yz \\
\frac{c_{4}}{2}z^{2}-\frac{c_{2}}{2}xy-\frac{c_{6}}{2}xz-\frac{%
c_{5}}{2}yz & \frac{c_{2}}{2}x^{2}+\frac{c_{3}}{2}%
z^{2} +c_{5}xz & \frac{c_{6}}{2}x^{2}-\frac{c_{5}}{2}xy -\frac{c_{4}}{2}xz-\frac{c_{3}}{2}yz \\
    \frac{c_{5}}{2}y^{2}-\frac{c_{6}}{2}xy-\frac{c_{1}}{2}xz-\frac{%
c_{4}}{2}yz & \frac{c_{6}}{2}x^{2}-\frac{c_{5}}{2}xy -\frac{c_{4}}{2}xz-\frac{c_{3}}{2}yz & \frac{c_{1}}{2}x^{2} +\frac{c_{3}}{2} y^{2}+c_{4}xy \\
  \end{array}
\right)
\end{equation*}
and the vector (\ref{eq.gencas.2}) becomes the general non-gradient KV
\[
K_{a}=
\left(
  \begin{array}{c}
   c_{7}y + c_{8}z \\
   -c_{7}x +c_{9}z \\
   -c_{8}x -c_{9}y \\
  \end{array}
\right).
\]

Then the constraint (\ref{eq.TKNq4}) implies that (since $K_{a}q^{a}=0$) $K=G(x,y,z)$ which when replaced in (\ref{eq.TKNq3}) gives (since $K_{ab}q^{b}=0$) $G_{,a}=0$. Hence $K=const \equiv0$.

The QFI $I=K_{ab}\dot{q}^{a}\dot{q}^{b} + K_{a}\dot{q}^{a}$ leads only to the three components $L_{i}$ of the angular momentum. We note that $I$ contains nine independent parameters each of them defining a FI: a) $c_{7}$, $c_{8}$, $c_{9}$ lead to the components $L_{1}= y\dot{z} - z\dot{y}$, $L_{2}= z\dot{x} - x\dot{z}$, $L_{3}= x\dot{y} - y\dot{x}$ of the angular momentum (LFIs); and b) $c_{1}$, $c_{2}$, $c_{3}$, $c_{4}$, $c_{5}$, $c_{6}$ lead to the products (QFIs depending on $L_{i}$) $L_{1}^{2}$, $L_{2}^{2}$, $L_{3}^{2}$, $L_{1}L_{2}$, $L_{1}L_{3}$ and $L_{2}L_{3}$.

We have the following result.

\begin{proposition} \label{pro.lfis}
The time-dependent generalized Kepler potential $V(t,q)= -\frac{\omega(t)}{r^{\nu}}$ for a general smooth function $\omega(t)$ admits only the LFIs of the angular momentum $L_{i}$. Independent QFIs in general do not exist, they are all quadratic combinations of $L_{i}$.
\end{proposition}

\subsection{$c_{A}=0$ where $A=1, 2, ..., 9$}

\label{sec.gen2}

In this case the conditions (\ref{eq.gen1.1}) imply that $a_{3}(t)= b_{0} + b_{1}t + b_{2}t^{2}$ and
\begin{equation}
\omega_{(\nu)}(t)= k\left(b_{0} + b_{1}t + b_{2}t^{2} \right)^{\frac{\nu-2}{2}} \label{eq.TKFI8}
\end{equation}
where $k, b_{0}, b_{1}, b_{2}$ are arbitrary constants and the index $(\nu)$ denotes the dependence of $\omega(t)$ on the value of $\nu$.

Since $c_{A}=0$ the quantities (\ref{eq.gencas.1}) and (\ref{eq.gencas.2}) become
\[
K_{ab}= a_{3}\delta_{ab}, \enskip K_{a}= -\dot{a}_{3}q_{a}.
\]

Substituting in the remaining constraints (\ref{eq.TKNq3}) and (\ref{eq.TKNq4}) we find
\[
K= b_{2}r^{2} - \frac{2k(b_{0}+b_{1}t+b_{2}t^{2})^{\nu/2}}{r^{\nu}}.
\]

The QFI is
\begin{equation}
J_{\nu}= (b_{0} + b_{1}t + b_{2}t^{2}) \left[ \frac{\dot{q}^{i}\dot{q}_{i}}{2} - \frac{k(b_{0} +b_{1}t + b_{2}t^{2})^{\frac{\nu-2}{2}}}{r^{\nu}} \right]-\frac{b_{1} + 2b_{2}t}{2} q^{i}\dot{q}_{i} +\frac{b_{2} r^{2}}{2}. \label{eq.TKFI7}
\end{equation}

We note that the resulting time-dependent generalized Kepler potential
\begin{equation}
V= -\frac{\omega_{\nu}(t)}{r^{\nu}}, \enskip \omega_{\nu}= k\left(b_{0} + b_{1}t + b_{2}t^{2} \right)^{\frac{\nu-2}{2}} \label{eq.gen1}
\end{equation}
is a  subcase of the Case III potential of \cite{Leach 1985} if we set the function
\[
U\left( \frac{r}{\phi} \right)= k_{1}\frac{r^{2}}{\phi^{2}} -\frac{k\phi^{\nu}}{r^{\nu}}
\]
with
\[
\phi= \sqrt{b_{0}+ b_{1}t +b_{2}t^{2}}, \enskip k_{1}= \frac{b_{0}b_{2}}{2} -\frac{b_{1}^{2}}{8}.
\]
Then the associated QFI (3.13) of \cite{Leach 1985} (for $K_{1}=K_{2}=0$) reduces to the QFI $J_{\nu}$.

For some values of $\nu$ we have the following results:

- $\nu=1$ (time-dependent Kepler potential).

The $\omega_{(1)}(t)= k\left(b_{0} + b_{1}t + b_{2}t^{2} \right)^{-1/2}$ and the QFI $J_{1}=E_{3}$ (see subsection \ref{sec.omega.3} below).

- $\nu=2$ (Newton-Cotes potential \cite{Ibragimov 1998}).

The $\omega_{(2)}=k=const$ and the QFI is
\begin{eqnarray*}
J_{2} &=& (b_{0} + b_{1}t + b_{2}t^{2})\left( \frac{\dot{q}^{i} \dot{q}_{i}}{2} - \frac{k}{r^{2}} \right) -\frac{b_{1} + 2b_{2} t}{2} q^{i}\dot{q}_{i} +\frac{b_{2}}{2} r^{2} \\
&=& b_{0}H_{2} -b_{1} I_{2} -b_{2}I_{1}.
\end{eqnarray*}
This expression contains the independent QFIs
\[
H_{2}= \frac{\dot{q}^{i}\dot{q}_{i}}{2} - \frac{k}{r^{2}}, \enskip I_{1}= -t^{2}H_{2} + tq^{i}\dot{q}_{i} -\frac{r^{2}}{2}, \enskip I_{2}= -tH_{2} +\frac{q^{i}\dot{q}_{i}}{2}
\]
where $H_{2}$ is the Hamiltonian of the system. These are the
FIs found in \cite{Tsamparlis 2020} (see also Table \ref{T1}) in the case of the autonomous generalized Kepler potential for $\nu=2$.

- $\nu=-2$ (time-dependent oscillator).

The $\omega_{(-2)}= k\left(b_{0} + b_{1}t + b_{2}t^{2} \right)^{-2}$ and the QFI is
\[
J_{-2}= (b_{0} + b_{1}t + b_{2}t^{2}) \left[ \frac{\dot{q}^{i}\dot{q}_{i}}{2} - \frac{k}{(b_{0} +b_{1}t + b_{2}t^{2})^{2}}r^{2} \right]-\frac{b_{1} + 2b_{2}t}{2} q^{i}\dot{q}_{i} +\frac{b_{2} r^{2}}{2}.
\]
This is the trace of the QFIs (\ref{eq.osc.FI3}) found below for $a_{3}(t)= b_{0} + b_{1}t + b_{2}t^{2}$. Substituting this $a_{3}(t)$ in (\ref{eq.osc.FI2}) and (\ref{eq.osc.FI3}) we find respectively that the $\omega= \omega_{(-2)}$ with constant $k=-\frac{1}{8}(b_{1}^{2} -4b_{2}b_{0} +2c_{0})$ and the QFIs are
\begin{equation}
I_{ij}= \Lambda_{ij}(a_{3}=b_{0}+b_{1}t+b_{2}t^{2}) = (b_{0}+b_{1}t+b_{2}t^{2}) \left( \dot{q}_{i}\dot{q}_{j} -2\omega q_{i}q_{j} \right) -(b_{1} +2b_{2}t)q_{(i}\dot{q}_{j)} +b_{2}q_{i}q_{j}. \label{eq.TKFI9}
\end{equation}
Therefore the trace $Tr[I_{ij}]= I_{11}+I_{22}+I_{33} =2J_{-2}$. Note that $r^{2}=q^{i}q_{i}$. \newline

We infer the following new general result which includes the time-dependent Kepler potential and the time-dependent oscillator as subcases.

\begin{proposition}[3d time-dependent generalized Kepler potentials which
admit FIs]
\label{oscillator}  For all functions $\omega(t)$ the time-dependent generalized Kepler potential $V(t,q)= -\frac{\omega(t)}{r^{\nu}}$ admits the LFIs of the angular momentum and QFIs which are products of the components of the angular momentum. However for the function $\omega(t)=\omega_{(\nu)}(t) =k\left(b_{0} + b_{1}t + b_{2}t^{2} \right)^{\frac{\nu-2}{2%
}}$ the resulting time-dependent generalized Kepler potential admits the additional QFI $J_{\nu}$ given by (\ref{eq.TKFI7}).
\end{proposition}

\section{The time-dependent Kepler potential}

\label{sec.timedep.Kepler}

In this case $\nu=1$ and conditions (\ref{eq.conTK1a}) - (\ref{eq.conTK1d}) give
\[
a_{16}=a_{17}=a_{18}=a_{19}=a_{20}=0, \enskip a_{5}=a_{8}, \enskip a_{2}=a_{12}, \enskip a_{3}=a_{9}=a_{13}, \enskip a_{11}=a_{15}
\]
\[
\ddot{a}_{2}= \ddot{a}_{5}= \ddot{a}_{11}=0
\]
\[
\sigma_{2}=c_{7}, \enskip \sigma_{3}= c_{8}, \enskip \tau_{3}= c_{9}.
\]
Then constraint (\ref{eq.TKNq5}) gives
\[
\dddot{a}_{3}=0, \enskip \sigma_{4}= \tau_{4}= \eta_{4}=0
\]
and
\[
a_{3}\omega^{2}=c_{10}, \enskip a_{2}\omega=c_{11}, \enskip a_{5}\omega=c_{12}, \enskip a_{11}\omega=c_{13}
\]
where $c_{10}, c_{11}, c_{12}, c_{13}$ are arbitrary constants.

Finally, we have
\begin{eqnarray*}
K_{11} &=&\frac{c_{2}}{2}y^{2}+\frac{c_{1}}{2}%
z^{2}+c_{6}yz+a_{5}y+a_{2}z+a_{3} \\
K_{12} &=&\frac{c_{4}}{2}z^{2}-\frac{c_{2}}{2}xy-\frac{c_{6}}{2}xz-\frac{%
c_{5}}{2}yz-\frac{a_{5}}{2}x-\frac{a_{11}}{2}y  \\
K_{13} &=&\frac{c_{5}}{2}y^{2}-\frac{c_{6}}{2}xy-\frac{c_{1}}{2}xz-\frac{%
c_{4}}{2}yz-\frac{a_{2}}{2}x-\frac{a_{11}}{2}z
\\
K_{22} &=&\frac{c_{2}}{2}x^{2}+\frac{c_{3}}{2}%
z^{2} +c_{5}xz+a_{11}x+a_{2}z+a_{3}  \\
K_{23} &=&\frac{c_{6}}{2}x^{2}-\frac{c_{5}}{2}xy -\frac{c_{4}}{2}xz-\frac{c_{3}}{2}yz -\frac{a_{2}}{2}y -\frac{a_{5}}{2}z \\
K_{33} &=&\frac{c_{1}}{2}x^{2}+\frac{c_{3}}{2}%
y^{2}+c_{4}xy+a_{11}x+a_{5}y+a_{3}
\end{eqnarray*}
and
\begin{eqnarray*}
K_{1} &=& \dot{a}_{11}y^{2} + \dot{a}_{11}z^{2} -\dot{a}_{5}xy -\dot{a}_{2}xz -\dot{a}_{3}x + c_{7}y + c_{8}z \\
K_{2} &=& \dot{a}_{5}x^{2} + \dot{a}_{5}z^{2} -\dot{a}_{11}xy -\dot{a}_{2}yz -c_{7}x -\dot{a}_{3}y + c_{9}z \\
K_{3} &=& \dot{a}_{2}x^{2} + \dot{a}_{2}y^{2} -\dot{a}_{11}xz -\dot{a}_{5}yz -c_{8}x -c_{9}y -\dot{a}_{3}z
\end{eqnarray*}
where
\begin{equation}
\ddot{a}_{2}= \ddot{a}_{5}= \ddot{a}_{11}=0, \enskip \dddot{a}_{3}=0, \enskip a_{3}\omega^{2}=c_{10}, \enskip a_{2}\omega=c_{11}, \enskip a_{5}\omega=c_{12}, \enskip a_{11}\omega=c_{13}. \label{Kepler.1}
\end{equation}

From the last conditions follow that in order QFIs to be admitted the
function $\omega(t)$ can have only three possible forms: \newline
- $\omega(t)$ a general function;\newline
- $\omega(t)= \omega_{2K}(t)= \frac{c_{11}}{b_{0} +b_{1}t}$ where $c_{11}b_{1}\neq0$; and\newline
- $\omega(t)= \omega_{3K}(t)=\frac{k}{(b_{0}+b_{1}t +b_{2}t^{2})^{1/2}}$ where $k\neq0$ and $b_{1}^{2} -4b_{2}b_{0} \neq0$.

This result confirms the results found previously in \cite{LeoTsampAndro 2017}, \cite{Katzin 1982}, \cite{Leach 1985}. We note that the time-dependent Kepler potential $V= -\frac{\omega_{2K}(t)}{r}$ is a subcase of the Case II potential of \cite{Leach 1985} for $\mu_{0}=c_{11}$ and $\phi=b_{0}+b_{1}t$, whereas the potential $V= -\frac{\omega_{3K}(t)}{r}$ is a subcase of the Case III potential of \cite{Leach 1985} (see subsection \ref{sec.gen2}).

In the following we discuss the cases for the special functions $\omega_{2K}(t)$ and $\omega_{3K}(t)$ because the case for a general function $\omega(t)$ reproduces the results of the subsection \ref{sec.gen1}.

\subsection{$\omega(t)= \omega_{2K}(t)= \frac{c_{11}}{b_{0} +b_{1}t}$, $c_{11}b_{1}\neq0$}

\label{sec.omega.2}

In that case conditions (\ref{Kepler.1}) give
\[
a_{2}=b_{0}+b_{1}t, \enskip a_{3}= \frac{c_{10}}{c_{11}^{2}} (b_{0}+b_{1}t)^{2}, \enskip a_{5}= \frac{c_{12}}{c_{11}} (b_{0}+b_{1}t), \enskip a_{11}= \frac{c_{13}}{c_{11}} (b_{0}+b_{1}t).
\]
Substituting the resulting vector $K_{a}$ and the KT $K_{ab}$ in (\ref{eq.TKNq4}) we find the solution
\[
K(q,t) = - \frac{2c_{10}b_{1}t}{c_{11}r} + G(q).
\]
Replacing this solution in the remaining constraint (\ref{eq.TKNq3}) we find
\[
G(x,y,z)= - \frac{2c_{10}b_{0}}{c_{11}r} - \frac{c_{13}x + c_{12}y + c_{11}z}{r} + \frac{c_{10}b_{1}^{2}}{c_{11}^{2}} r^{2}.
\]
Therefore
\[
K(x,y,z,t)= \frac{c_{10}b_{1}^{2} r^{2}}{c_{11}^{2}} -\frac{2c_{10}(b_{0}+b_{1}t)}{c_{11}r} -\frac{c_{13}x +c_{12}y + c_{11}z}{r}.
\]

The QFI is
\begin{eqnarray*}
I&=& \frac{c_{3}}{2} L_{1}^{2} +\frac{c_{1}}{2} L_{2}^{2} +\frac{c_{2}}{2} L_{3}^{2} -c_{4}L_{1}L_{2} -c_{5}L_{1}L_{3} -c_{6}L_{2}L_{3} -c_{9}L_{1} + c_{8}L_{2} - c_{7}L_{3} + \frac{2c_{10}}{c_{11}^{2}}E_{2} + \\
&& + \frac{c_{13}}{c_{11}} A_{1}+ \frac{c_{12}}{c_{11}} A_{2} + A_{3}
\end{eqnarray*}
where $\omega_{2K}(t)=\frac{c_{11}}{b_{0}+b_{1}t}$ and
\begin{eqnarray}
L_{i} &\equiv& q_{i+1} \dot{q}_{i+2} - q_{i+2} \dot{q}_{i+1} \label{eq.TKFI1} \\
E_{2} &\equiv& (b_{0}+b_{1}t)^{2} \left[ \frac{\dot{q}^{i} \dot{q}_{i}}{2} -\frac{c_{11}}{r(b_{0}+b_{1}t)} \right] -b_{1}(b_{0}+b_{1}t) q^{i}\dot{q}_{i} + \frac{b_{1}^{2}r^{2}}{2} \label{eq.TKFI1a} \\
\tilde{R}_{i} &\equiv& (\dot{q}^{j}\dot{q}_{j})q_{i} - (\dot{q}^{j}q_{j})\dot{q}_{i} - \frac{c_{11}}{r(b_{0}+b_{1}t)} q_{i} \label{eq.TKFI1b} \\
A_{i} &\equiv& (b_{0}+b_{1}t) \tilde{R}_{i} + b_{1}\left( q_{i+2}L_{i+1} - q_{i+1}L_{i+2} \right). \label{eq.TKFI1c}
\end{eqnarray}
We note that $i=1,2,3$, $q_{i}=(x,y,z)$ and $q_{i}\equiv q_{i+3k}$ for all $k \in \mathbb{N}$, that is
\[
x=q_{1}=q_{4}=q_{7}=..., \enskip y=q_{2}=q_{5}=q_{8}=..., \enskip z=q_{3}=q_{6}=q_{9}.
\]

The QFI $I$ contains the already found LFIs $L_{i}$ of the angular momentum; the QFI $E_{2}$ which for $b_{1}=0$ reduces to the Hamiltonian of the Kepler potential $V= -\frac{c_{11}}{b_{0}r}$; and the QFIs $A_{i}$ which may be considered as a generalization of the Runge-Lenz vector $R_{i}\left( k=\frac{c_{11}}{b_{0}} \right)$ for time-dependence $\omega_{2K}(t)= \frac{c_{11}}{b_{0}+b_{1}t}$. Indeed we have $A_{i}(b_{1}=0)= b_{0}R_{i}\left( k=\frac{c_{11}}{b_{0}} \right)$.

The expressions (\ref{eq.TKFI1a}) - (\ref{eq.TKFI1c}) are written compactly as follows
\begin{eqnarray}
E_{2} &\equiv& c_{11}^{2} \left[ \frac{1}{\omega_{2K}^{2}} \left( \frac{\dot{q}^{i}\dot{q}_{i}}{2} -\frac{\omega_{2K}}{r} \right) -\frac{1}{2} \frac{d}{dt}\left(\frac{1}{\omega_{2K}}\right)^{2} q^{i}\dot{q}_{i} + \frac{d^{2}}{dt^{2}} \left( \frac{1}{\omega_{2K}} \right)^{2} \frac{r^{2}}{4} \right] \label{eq.TKFI3a} \\
\tilde{R}_{i} &\equiv& (\dot{q}^{j}\dot{q}_{j})q_{i} - (\dot{q}^{j}q_{j})\dot{q}_{i} - \frac{\omega_{2K}}{r}q_{i} \label{eq.TKFI3b} \\
A_{i} &\equiv& c_{11} \left[ \frac{1}{\omega_{2K}} \tilde{R}_{i} - \frac{(\ln \omega_{2K})^{\cdot}}{\omega_{2K}}\left( q_{i+2}L_{i+1} - q_{i+1}L_{i+2} \right) \right] \label{eq.TKFI3c}
\end{eqnarray}
where $\omega_{2K}(t)= \frac{c_{11}}{b_{0}+b_{1}t}$.

We remark that only five of the seven FIs $E_{2}$, $L_{i}$, $A_{i}$ are functionally independent because they are related as follows
\begin{equation}
\mathbf{A} \cdot \mathbf{L} = 0, \enskip 2E_{2}\mathbf{L}^{2} + c_{11}^{2} = \mathbf{A}^{2}. \label{eq.TKFI4}
\end{equation}
For $b_{1}=0$, $b_{0}\neq0$ we have $\omega_{2K}=\frac{c_{11}}{b_{0}}\equiv k=const$, $E_{2}=b_{0}^{2}H$, $\tilde{R}_{i}=R_{i}$ and $A_{i}= b_{0}R_{i}$ where $H$ is the Hamiltonian and $R_{i}$ the Runge-Lenz vector for the Kepler potential $V=-\frac{k}{r}$. Then, as expected, equation (\ref{eq.TKFI4}) reduces to the well-known relation
\[
2H\mathbf{L}^{2} + k^{2} = \mathbf{R}^{2}.
\]

\subsection{$\omega(t)= \omega_{3K}(t)= \frac{k}{(b_{0}+b_{1}t +b_{2}t^{2})^{1/2}}$, $k\neq0$, $b_{1}^{2} -4b_{2}b_{0}\neq0$}

\label{sec.omega.3}

In that case conditions (\ref{Kepler.1}) give\footnote{Observe that if $b_{1}^{2}-4b_{2}b_{0}=0$ this case reduces to the case of the subsection \ref{sec.omega.2} because equation $b_{0}+b_{1}t+b_{2}t^{2}=0$ has a double root $t_{0}$ and can be factored in the form $b_{2}(t-t_{0})^{2}$.}
\[
a_{2}=a_{5}=a_{11}=0, \enskip c_{11}= c_{12}= c_{13}=0, \enskip a_{3}= \frac{c_{10}}{k^{2}}(b_{0}+b_{1}t+b_{2}t^{2}).
\]

Substituting the $K_{a}$ and $K_{ab}$ of that case in (\ref{eq.TKNq4}) we find the solution
\[
K(q,t)= -\frac{2c_{10}}{r\omega_{3K}} + G(q).
\]
When this solution is introduced in the remaining constraint (\ref{eq.TKNq3}) gives $G(x,y,z)= \frac{b_{2}c_{10}}{k^{2}}r^{2}$. Therefore
\[
K(x,y,z,t) = \frac{b_{2}c_{10}}{k^{2}}r^{2} -\frac{2c_{10}}{r\omega_{3K}}.
\]

The QFI is
\begin{equation*}
I = \frac{c_{3}}{2} L_{1}^{2} +\frac{c_{1}}{2} L_{2}^{2} +\frac{c_{2}}{2} L_{3}^{2} -c_{4}L_{1}L_{2} -c_{5}L_{1}L_{3} -c_{6}L_{2}L_{3} -c_{9}L_{1} + c_{8}L_{2} - c_{7}L_{3} + \frac{2c_{10}}{k^{2}} E_{3}
\end{equation*}
where
\begin{equation}
E_{3} \equiv (b_{0}+b_{1}t+b_{2}t^{2}) \left[ \frac{\dot{q}^{i}\dot{q}_{i}}{2} -\frac{k}{r(b_{0}+b_{1}t +b_{2}t^{2})^{1/2}} \right] - \frac{b_{1}+2b_{2}t}{2} q^{i}\dot{q}_{i} + \frac{b_{2}r^{2}}{2} \label{eq.TKFI5a}
\end{equation}
is the only new independent QFI. This QFI is written equivalently \begin{equation}
E_{3} = k^{2} \left[ \frac{1}{\omega_{3K}^{2}} \left( \frac{\dot{q}^{i}\dot{q}_{i}}{2} -\frac{\omega_{3K}}{r} \right) -\frac{1}{2} \frac{d}{dt}\left(\frac{1}{\omega_{3K}}\right)^{2} q^{i}\dot{q}_{i} + \frac{d^{2}}{dt^{2}} \left( \frac{1}{\omega_{3K}} \right)^{2} \frac{r^{2}}{4} \right]. \label{eq.TKFI5b}
\end{equation}
For $b_{1}=b_{2}=0$, $E_{2}$ reduces to the well-known Hamiltonian of the time-independent Kepler potential.

We note also that the QFIs (\ref{eq.TKFI1a}), (\ref{eq.TKFI5a})
can be written compactly as (see eq. (2.86) in \cite{Katzin 1982})
\begin{equation}
E_{\mu} = k^{2} \left[ \frac{1}{\omega_{\mu K}^{2}} \left( \frac{\dot{q}^{i}\dot{q}_{i}}{2} -\frac{\omega_{\mu K}}{r} \right) -\frac{1}{2} \frac{d}{dt}\left(\frac{1}{\omega_{\mu K}}\right)^{2} q^{i}\dot{q}_{i} + \frac{d^{2}}{dt^{2}} \left( \frac{1}{\omega_{\mu K}} \right)^{2} \frac{r^{2}}{4} \right] \label{eq.TKFI6com}
\end{equation}
where $\mu=2,3$, $\omega_{2K}(t)= \frac{k}{b_{0}+b_{1}t}$ and $\omega_{3K}(t)= \frac{k}{(b_{0}+b_{1}t+b_{2}t^{2})^{1/2}}$.

\begin{proposition}[Time-dependent Kepler potentials which admit additional FIs \cite{Katzin 1982}]
\label{kepler} The time-dependent Kepler
potential $V(t,q)=-\frac{\omega (t)}{r}$ for the function $\omega _{2K}(t)=\frac{c_{11}}{b_{0}+b_{1}t%
}$ , $c_{11}b_{1}\neq 0$ and the function $\omega_{3K}(t)=\frac{k}{%
(b_{0}+b_{1}t+b_{2}t^{2})^{1/2}}$ where $k\neq 0$ and $b_{1}^{2}-4b_{2}b_{0}%
\neq 0$ admits additional QFIs given by (\ref{eq.TKFI1a}), (\ref{eq.TKFI1c}) and (\ref{eq.TKFI5a}) respectively.
\end{proposition}

\section{The 3d time-dependent oscillator}

In this case $\nu =-2$ and conditions (\ref{eq.conTK1a}) - (\ref{eq.conTK1d}) give
\[
a_{2}=a_{5}=a_{8}=a_{11}=a_{12}=a_{15}=a_{16}=a_{18}=0
\]
and
\begin{equation}
\enskip \dot{\sigma}_{2}= -\ddot{a}_{17}, \enskip \dot{\sigma}_{3}= -\ddot{a}_{19}, \enskip \dot{\tau}_{3}= -\ddot{a}_{20}. \label{eq.osc.pde1}
\end{equation}
Then the constraint (\ref{eq.TKNq5}) implies that
\begin{equation}
\ddot{\sigma}_{4} - 2\omega\sigma_{4}=0, \enskip \ddot{\tau}_{4} - 2\omega\tau_{4}=0, \enskip \ddot{\eta}_{4} - 2\omega\eta_{4}=0, \label{eq.osc.pde1a}
\end{equation}
\begin{equation}
\dddot{a}_{3} - 8\omega\dot{a}_{3} -4\dot{\omega}a_{3}=0, \enskip \dddot{a}_{9} - 8\omega\dot{a}_{9} -4\dot{\omega}a_{9}=0, \enskip \dddot{a}_{13} - 8\omega\dot{a}_{13} -4\dot{\omega}a_{13}=0, \label{eq.osc.pde1b}
\end{equation}
\begin{equation}
\dddot{a}_{17} - 8\omega\dot{a}_{17} -4\dot{\omega}a_{17}=0,
\enskip \dddot{a}_{19} - 8\omega\dot{a}_{19} -4\dot{\omega}a_{19}=0, \enskip \dddot{a}_{20} - 8\omega\dot{a}_{20} -4\dot{\omega}a_{20}=0. \label{eq.osc.pde1c}
\end{equation}

Therefore
\begin{eqnarray}
K_{11} &=&\frac{c_{2}}{2}y^{2}+\frac{c_{1}}{2}%
z^{2}+c_{6}yz+a_{3} \notag \\
K_{12} &=&\frac{c_{4}}{2}z^{2}-\frac{c_{2}}{2}xy-\frac{c_{6}}{2}xz-\frac{%
c_{5}}{2}yz+a_{17} \notag \\
K_{13} &=&\frac{c_{5}}{2}y^{2}-\frac{c_{6}}{2}xy-\frac{c_{1}}{2}xz-\frac{%
c_{4}}{2}yz+a_{19} \notag
\\
K_{22} &=&\frac{c_{2}}{2}x^{2}+\frac{c_{3}}{2}%
z^{2} +c_{5}xz+a_{13} \label{eq.osc.pde2} \\
K_{23} &=&\frac{c_{6}}{2}x^{2}-\frac{c_{5}}{2}xy -\frac{c_{4}}{2}xz-\frac{c_{3}}{2}yz +a_{20} \notag \\
K_{33} &=&\frac{c_{1}}{2}x^{2}+\frac{c_{3}}{2}
y^{2}+c_{4}xy +a_{9} \notag
\end{eqnarray}
and
\begin{eqnarray}
K_{1} &=& -\dot{a}_{3}x + \sigma_{2}y + \sigma_{3}z + \sigma_{4} \notag \\
K_{2} &=& -(\sigma_{2}+2\dot{a}_{17})x -\dot{a}_{13}y + \tau_{3}z + \tau_{4} \label{eq.osc.pde3} \\
K_{3} &=& -(\sigma_{3}+2\dot{a}_{19})x -(\tau_{3}+2\dot{a}_{20})y -\dot{a}_{9}z + \eta_{4}. \notag
\end{eqnarray}

Before we proceed with considering various subcases it is important that we discuss the ordinary differential equations (ODEs) (\ref{eq.osc.pde1b}) and (\ref{eq.osc.pde1c}).

\subsection{The Lewis invariant}

\label{Note 1}

Equations of the form
\begin{equation}
\dddot{a} - 8\omega\dot{a} -4\dot{\omega}a=0 \label{eq.TKNq10}
\end{equation}
where $a=a(t)$ can be written as follows
\begin{equation}
a\ddot{a} - \frac{1}{2}\dot{a}^{2} -4\omega a^{2} = c_{0}=const. \label{eq.TKNq11}
\end{equation}
By putting $a=-\rho^{2}$ where $\rho=\rho(t)$ equation (\ref{eq.TKNq11}) becomes
\begin{equation}
\ddot{\rho} -2\omega \rho - \frac{c_{0}}{2\rho^{3}}=0. \label{eq.TKNq12}
\end{equation}
For $2\omega(t)=-\psi^{2}(t)$ equation (\ref{eq.TKNq12}) is written
\begin{equation}
\ddot{\rho} + \psi^{2}\rho - \frac{c_{0}}{2\rho^{3}}=0. \label{eq.TKNq13}
\end{equation}
Equation (\ref{eq.TKNq13}) is the auxiliary equation (see \cite{Katzin 1977}, \cite{Leach 1991}, \cite{Tsamparlis 2012}) that should be introduced in order to derive the Lewis invariant for the one-dimensional (1d) time-dependent oscillator
\begin{equation}
\ddot{x} + \psi^{2}x =0. \label{eq.TKNq14}
\end{equation}
By eliminating the $\psi^{2}$ using (\ref{eq.TKNq14}) and multiplying with the factor $x\dot{\rho}-\rho\dot{x}$ equation (\ref{eq.TKNq13}) gives
\[
\ddot{\rho} - \frac{\rho}{x}\ddot{x} - \frac{c_{0}}{2\rho^{3}}=0 \implies
\left[ \frac{1}{2}\left( x\dot{\rho} - \rho\dot{x} \right)^{2} + \frac{c_{0}}{4}\left(\frac{x}{\rho}\right)^{2} \right]^{\cdot} =0 \implies
\]
\begin{equation}
I \equiv \frac{1}{2}\left( x\dot{\rho} - \rho\dot{x} \right)^{2} +\frac{c_{0}}{4}\left(\frac{x}{\rho}\right)^{2} =const \label{eq.TKNq15}
\end{equation}
which is the well-known Lewis invariant for the 1d time-dependent harmonic oscillator or, equivalently, a FI for the two-dimensional (2d) time-dependent system with equations of motion (\ref{eq.TKNq13}) and (\ref{eq.TKNq14}).

\subsection{The system of equations (\ref{eq.osc.pde1}) - (\ref{eq.osc.pde1c})}

The conditions (\ref{eq.osc.pde1a}) are not involved into the conditions (\ref{eq.osc.pde1}), (\ref{eq.osc.pde1b}) and (\ref{eq.osc.pde1c}). This means that the parameters $\sigma_{4}, \tau_{4}, \eta_{4}$ give different independent FIs from the remaining parameters $a_{3}, a_{9}, a_{13}, a_{17}, a_{19}, a_{20}$. Therefore without loss of generality they can be treated separately. This leads to the following two cases.

\subsubsection{$a_{3}\neq0$, $\sigma_{4} =\tau_{4} =\eta_{4} =0$}

\label{2.3}

Because the ODEs (\ref{eq.osc.pde1b}), (\ref{eq.osc.pde1c}) are independent (i.e. each one leads to a different FI) and are of the same form without loss of generality we assume
\[
a_{9}=k_{1}a_{3}, \enskip a_{13}=k_{2}a_{3}, \enskip a_{17}=k_{3}a_{3}, \enskip a_{19}= k_{4}a_{3}, \enskip a_{20}= k_{5}a_{3}
\]
where $k_{1}, k_{2}, k_{3}, k_{4}, k_{5}$ are arbitrary constants.

From the discussion of subsection \ref{Note 1} and the assumption $a_{3}\neq0$ condition (\ref{eq.osc.pde1b}) concerning $a_{3}(t)$ becomes (see eq. (9.2) in \cite{Katzin 1977})
\begin{equation}
\dddot{a}_{3} - 8\omega\dot{a}_{3} -4\dot{\omega}a_{3}=0 \implies a_{3}\ddot{a}_{3} - \frac{1}{2}\dot{a}_{3}^{2} -4\omega a_{3}^{2} = c_{0} \implies \omega(t)= \frac{\ddot{a}_{3}}{4a_{3}} - \frac{1}{8} \left(\frac{\dot{a}_{3}}{a_{3}}\right)^{2} -\frac{c_{0}}{4a_{3}^{2}} \label{eq.osc.FI2}
\end{equation}
where $c_{0}$ is an arbitrary constant and $a_{3}(t)$ is an arbitrary non-zero function.

Moreover, conditions (\ref{eq.osc.pde1}) become
\[
\sigma_{2}=-\dot{a}_{17}, \enskip \sigma_{3}= -\dot{a}_{19}, \enskip \tau_{3}= -\dot{a}_{20}
\]
because any additional constant (in general $\sigma_{2}= -\dot{a}_{17} +m_{1}$ where $m_{1}$ is a constant) leads to the usual LFIs of the angular momentum.

Then the KT (\ref{eq.osc.pde2}) and the vector (\ref{eq.osc.pde3}) become\footnote{We set $c_{1}= ... = c_{6} =0$ because they generate the already found FIs of the angular momentum.}
\begin{equation*}
K_{ab}=a_{3}\left(
\begin{array}{ccc}
1 & k_{3} & k_{4} \\
k_{3} & k_{2} & k_{5} \\
k_{4} & k_{5} & k_{1}%
\end{array}%
\right) ,\enskip K_{a}=-\dot{a}_{3}\left(
\begin{array}{c}
x+k_{3}y+k_{4}z \\
k_{3}x+k_{2}y+k_{5}z \\
k_{4}x+k_{5}y+k_{1}z%
\end{array}%
\right).
\end{equation*}

Substituting in the constraints (\ref{eq.TKNq3}) and (\ref{eq.TKNq4}) we find
\begin{equation*}
K=\frac{\dot{a}_{3}^{2}+2c_{0}}{4a_{3}}\left(
x^{2}+k_{2}y^{2}+k_{1}z^{2}+2k_{3}xy+2k_{4}xz+2k_{5}yz\right) .
\end{equation*}%
Using equation (\ref{eq.osc.FI2}) we can write $\frac{\dot{a}_{3}^{2}+2c_{0}%
}{4a_{3}}=\frac{\ddot{a}_{3}}{2}-2\omega a_{3}$.

The QFI is
\begin{eqnarray*}
I &=& a_{3}\left( \dot{x}^{2} + k_{2}\dot{y}^{2} +k_{1}\dot{z}^{2} +
2k_{3}\dot{x}\dot{y} + 2k_{4}\dot{x}\dot{z} + 2k_{5}\dot{y}\dot{z} \right) -%
\dot{a}_{3}(x +k_{3}y +k_{4}z)\dot{x} - \\
&& -\dot{a}_{3}(k_{3}x +k_{2}y+ k_{5}z)\dot{y} -\dot{a}_{3}(k_{4}x + k_{5}y
+k_{1}z)\dot{z}+ \\
&& + \left( \frac{\ddot{a}_{3}}{2} -2\omega a_{3} \right) \left( x^{2} +
k_{2}y^{2} +k_{1}z^{2} +2k_{3}xy +2k_{4}xz +2k_{5}yz \right).
\end{eqnarray*}
This expression contains six QFIs which are the components of the symmetric tensor (see eqs. (1.4) and (6.24) in \cite{Katzin 1977})
\begin{equation}
\Lambda_{ij}= a_{3} \left( \dot{q}_{i}\dot{q}_{j} -2\omega q_{i} q_{j} \right) -\dot{a}_{3} q_{(i}\dot{q}_{j)} +\frac{\ddot{a}_{3}}{2} q_{i}q_{j}. \label{eq.osc.FI3}
\end{equation}
This tensor for $a_{3}=const\neq0$ reduces to the Jauch-Hill-Fradkin tensor $B_{ij}$ for $\omega= -\frac{c_{0}}{4a_{3}^{2}} =const$.

If we make the transformation (see subsection \ref{Note 1}) $a_{3}(t)=-\rho^{2}(t)$ and $2\omega(t)=-\psi^{2}(t)$, equation (\ref{eq.TGKep.1a}) becomes
\begin{equation}
\ddot{q}^{a} -2\omega q^{a} =0 \implies \ddot{q}^{a} + \psi^{2}q^{a} =0 \label{eq.osc.FI3a}
\end{equation}
and the QFIs (\ref{eq.osc.FI3}) give
\begin{equation}
\Lambda_{ij}= -\left(\rho\dot{q}_{i} - \dot{\rho}q_{i}\right) \left(\rho\dot{q}_{j} - \dot{\rho}q_{j}\right) -\frac{c_{0}}{2} \rho^{-2}q_{i}q_{j} \label{eq.osc.FI3b}
\end{equation}
where the condition (\ref{eq.osc.FI2}) takes the form (\ref{eq.TKNq13}).

The symmetric tensor (\ref{eq.osc.FI3b}) may be thought of as a 3d generalization of the 1d Lewis invariant (\ref{eq.TKNq15}). Moreover equation (\ref{eq.osc.FI3b}) coincides with eq. (8) in \cite{Gunther 1977} and eq. (1.4) in \cite{Katzin 1977} when $c_{0}=2$.

\subsubsection{$a_{3}=a_{9}=a_{13}=a_{17}=a_{19}=a_{20}=0$, $\sigma_{4}\neq0$}

\label{2.4}

In this case the conditions (\ref{eq.osc.pde1b}), (\ref{eq.osc.pde1c}) vanish identically; and the conditions (\ref{eq.osc.pde1}) imply that $\sigma_{2}=c_{7}$, $\sigma_{3}=c_{8}$ and $\tau_{3}=c_{9}$.

Since the remaining ODEs (\ref{eq.osc.pde1a}) are all independent (i.e. each one generates an independent FI) and of the same form without loss of generality we assume
\[
\tau_{4}=k_{1}\sigma_{4}, \enskip \eta_{4}=k_{2}\sigma_{4}
\]
where $k_{1}, k_{2}$ are arbitrary constants.

From (\ref{eq.osc.pde1a}) for $\sigma_{4}\neq0$ we get
\begin{equation}
\omega(t)= \frac{\ddot{\sigma}_{4}}{2\sigma_{4}}. \label{eq.osc.FI4}
\end{equation}

The parameters $c_{A}$ where $A=1,2,...,9$ produce the FIs of the angular momentum and we fix them to zero. Therefore
\[
K_{ab}=0, \enskip K_{a}=\sigma_{4} \left( 1, k_{1}, k_{2} \right).
\]
Substituting in the remaining constraints (\ref{eq.TKNq3}) and (\ref{eq.TKNq4}) we find
\[
K= -\dot{\sigma}_{4} \left(x + k_{1}y +k_{2}z \right).
\]

The QFI is
\[
I= \sigma_{4}\dot{x} -\dot{\sigma}_{4}x + k_{1} \left( \sigma_{4}\dot{y} -\dot{\sigma}_{4} y \right) + k_{2} \left( \sigma_{4}\dot{z} -\dot{\sigma}_{4}z \right)
\]
which contains the irreducible LFIs (see eq. (6.25) in \cite{Katzin 1977})
\begin{equation}
I_{4i}= f\dot{q}_{i} -\dot{f}q_{i} \label{eq.osc.FI5}
\end{equation}
where $f(t)$ is an arbitrary non-zero function satisfying (\ref{eq.osc.FI4}). We note that the LFIs (\ref{eq.osc.FI5}) can be derived directly from the equations of motion for $\omega(t)= \frac{\ddot{f}}{2f}$.
\bigskip

From the above two cases we arrive at the following conclusion.

\begin{proposition}[3d time-dependent oscillators which admit additional FIs]
\label{oscillator} For the function $\omega(t)= \frac{\ddot{a}_{3}}{4a_{3}} - \frac{1}{8} \left(\frac{\dot{a}_{3}}{a_{3}}\right)^{2} -\frac{c_{0}}{4a^{2}_{3}}$ where $a_{3}(t)\neq0$, $c_{0}$ is an arbitrary constant and the function $\omega(t)= \frac{\ddot{f}}{2f}$ where $f(t)\neq0$ the resulting 3d time-dependent oscillator $V(t,q)= -\omega(t)r^{2}$ admits the QFIs (\ref{eq.osc.FI3}) and the LFIs (\ref{eq.osc.FI5}) respectively.
\end{proposition}

\section{A special class of time-dependent oscillators}

\label{sec.discussion.GK}

In proposition \ref{oscillator} it has been shown that the time-dependent oscillator ($\nu=-2$) for the frequency
\begin{equation}
\omega_{1O}(t)=\frac{\ddot{f}}{4f(t)}-\frac{1}{8}\left( \frac{\dot{f}}{f}%
\right) ^{2}-\frac{c_{0}}{4f^{2}}  \label{eq.disosc0.a}
\end{equation}%
where $f(t)$ is an arbitrary non-zero function admits  the six QFIs
\begin{equation}
\Lambda _{ij}=f(t)\left( \dot{q}_{i}\dot{q}_{j}-2\omega q_{i}q_{j}\right) -%
\dot{f}q_{(i}\dot{q}_{j)}+\frac{\ddot{f}}{2}q_{i}q_{j}  \label{eq.disosc0.a1}
\end{equation}%
and for the frequency
\begin{equation}
\omega_{2O}(t)=\frac{\ddot{g}}{2g(t)}  \label{eq.disosc0.b}
\end{equation}
where $g(t)$ is an arbitrary non-zero function admits the three
LFIs
\begin{equation}
I_{4i}=g(t)\dot{q}_{i}-\dot{g}q_{i}.  \label{eq.disosc0.b1}
\end{equation}

We consider the class of the 3d time-dependent oscillators for which $\omega_{1O}(t)= \omega_{2O}(t)$. These oscillators admit both the six QFIs $\Lambda _{ij}$ and the three LFIs $I_{4i}$.

The condition $\omega _{1O}(t) =\omega _{2O}(t)$ relates the functions $f(t), g(t)$ as follows
\begin{equation}
\omega_{3O}(t)=\frac{\ddot{f}}{4f(t)}-\frac{1}{8}\left( \frac{\dot{f}}{f}%
\right) ^{2}-\frac{c_{0}}{4f^{2}}=\frac{\ddot{g}}{2g(t)}.  \label{eq.disosc1}
\end{equation}

It can be easily proved that
\begin{equation}
g=f^{1/2} \cos\theta, \enskip \dot{\theta}= \left( \frac{c_{0}}{2} \right)^{1/2} f^{-1} \implies  \theta(t)= \left( \frac{c_{0}}{2} \right)^{1/2} \int \frac{dt}{f(t)} \label{eq.disosc2a}
\end{equation}
and
\begin{equation}
g=f^{1/2} \sin\theta, \enskip \dot{\theta}= \left( \frac{c_{0}}{2} \right)^{1/2} f^{-1} \implies  \theta(t)= \left( \frac{c_{0}}{2} \right)^{1/2} \int \frac{dt}{f(t)} \label{eq.disosc2b}
\end{equation}
satisfy the requirement (\ref{eq.disosc1}) for any non-zero function $f(t)$. In other words all the time-dependent oscillators with frequency
\begin{equation}
\omega_{3O}(t)=\frac{\ddot{f}}{4f(t)} - \frac{1}{8} \left(\frac{\dot{f}}{f}\right)^{2} -\frac{c_{0}}{4f^{2}} \label{eq.disosc3}
\end{equation}
admit the six QFIs
\begin{equation}
\Lambda_{ij}= f(t) \left( \dot{q}_{i}\dot{q}_{j} -2\omega q_{i} q_{j} \right) -\dot{f} q_{(i}\dot{q}_{j)} +\frac{\ddot{f}}{2} q_{i}q_{j} \label{eq.disosc4a}
\end{equation}
and the six LFIs
\begin{eqnarray}
I_{41i}&=& \left( \frac{c_{0}}{2} \right)^{1/2} f^{-1/2}q_{i} \sin\theta + \left( f^{1/2} \dot{q}_{i} - \frac{\dot{f}}{2} f^{-1/2}q_{i} \right) \cos\theta \label{eq.disosc4b} \\
I_{42i}&=& -\left( \frac{c_{0}}{2} \right)^{1/2} f^{-1/2}q_{i} \cos\theta + \left( f^{1/2} \dot{q}_{i} - \frac{\dot{f}}{2} f^{-1/2}q_{i} \right) \sin\theta. \label{eq.disosc4c}
\end{eqnarray}
These are the LFIs $J^{k}_{3}$, $J^{k}_{4}$ derived in eqs. (44), (45) in \cite{Prince 1980} using Noether point symmetries and Noether's theorem.

We note that
\begin{equation}
\frac{dI_{42i}}{d\theta }=I_{41i}  \label{eq.disosc4c.1}
\end{equation}
and
\begin{equation}
\Lambda_{ij} = I_{41i}I_{41j} + I_{42i}I_{42j}. \label{eq.disosc5}
\end{equation}

Next we consider the LFIs of the angular momentum $L_{i}= q_{i+1} \dot{q}_{i+2} - q_{i+2} \dot{q}_{i+1}$ which can be expressed equivalently as components of the totally antisymmetric tensor
\begin{equation}
L_{ij}= q_{i}\dot{q}_{j} - q_{j}\dot{q}_{i} =\varepsilon_{ijk} L^{k} \label{eq.disosc6}
\end{equation}
where $\varepsilon_{ijk}$ is the 3d Levi-Civita symbol and $L^{i}= L_{i}$ since the kinetic metric $\gamma_{ij} =\delta_{ij}$. Then (see eq. (51) in \cite{Prince 1980})
\begin{equation}
L_{ij}= \left( \frac{2}{c_{0}} \right)^{1/2} \left( I_{41i}I_{42j} - I_{41j}I_{42i} \right). \label{eq.disosc7}
\end{equation}

\begin{proposition} \label{pro.indLFIs}
For the class of 3d time-dependent oscillators with potential $V(t,q)=-\omega(t)r^{2}$ where $\omega(t)$ is defined in terms of an arbitrary non-zero (smooth) function $f(t)$ as in (\ref{eq.disosc3}), the only independent FIs are the LFIs $I_{41i}, I_{42i}$.
\end{proposition}

In order to recover the results of \cite{Prince 1980}, we assume a time-dependent oscillator with $\omega_{3O}(t)$ given by (\ref{eq.disosc3}) and we write the non-zero function $f(t)$ in the form $f(t)= \rho^{2}(t)$. Then equation (\ref{eq.disosc3}) becomes
\begin{equation}
\omega_{3O}(t)= \frac{\ddot{\rho}}{2\rho} - \frac{c_{0}}{4\rho^{4}}. \label{eq.disosc8}
\end{equation}
The relations (\ref{eq.disosc2a}), (\ref{eq.disosc2b}) become
\begin{equation}
g=\rho \cos\theta, \enskip \dot{\theta}= \left( \frac{c_{0}}{2} \right)^{1/2} \rho^{-2} \implies  \theta(t)= \left( \frac{c_{0}}{2} \right)^{1/2} \int \frac{dt}{\rho^{2}} \label{eq.disosc8a}
\end{equation}
\begin{equation}
g=\rho \sin\theta, \enskip \dot{\theta}= \left( \frac{c_{0}}{2} \right)^{1/2} \rho^{-2} \implies  \theta(t)= \left( \frac{c_{0}}{2} \right)^{1/2} \int \frac{dt}{\rho^{2}} \label{eq.disosc8b}
\end{equation}
and the LFIs (\ref{eq.disosc4b}), (\ref{eq.disosc4c}) take the form
\begin{eqnarray}
I_{41i} &=& \left( \frac{c_{0}}{2} \right)^{1/2} \rho^{-1}q_{i} \sin\theta + \left( \rho \dot{q}_{i} - \dot{\rho}q_{i} \right) \cos\theta \label{eq.disosc9a} \\
I_{42i}&=& -\left( \frac{c_{0}}{2} \right)^{1/2} \rho^{-1}q_{i} \cos\theta + \left( \rho \dot{q}_{i} - \dot{\rho}q_{i} \right) \sin\theta. \label{eq.disosc9b}
\end{eqnarray}
These latter expressions for $c_{0}=2$ coincide with the independent LFIs (44) and (45) found in \cite{Prince 1980}.

Finally we note that if we consider in this special class of oscillators the simple case $f=1$, we find $\omega_{3O}(t)= const= -\frac{c_{0}}{4} \equiv k$ which is the 3d autonomous oscillator (for $k<0$). Then it can be shown that the exponential LFIs $I_{3i\pm}$ (see Table \ref{T1}) found in \cite{Tsamparlis 2020} can be written in terms of $I_{41i}, I_{42j}$. Indeed we have  $I_{3i\pm}(k>0) = I_{41i} \mp iI_{42i}$ and $I_{3i\pm}(k<0)= I_{41i} \pm iI_{42i}$.

\section{Collection of results}

\label{Table 2}

We collect the results concerning the time-dependent generalized Kepler potential for all values of $\nu$ in Table \ref{T2}. We note that for $\nu=-2,1,2$ the dynamical system is the time-dependent 3d oscillator, the time-dependent Kepler potential and the Newton-Cotes potential respectively. Concerning notation we have $q^{i}=(x,y,z)$, $q_{i}\equiv q_{i+3k}$ for all $k \in \mathbb{N}$ and $%
\tilde{R}_{i}= (\dot{q}^{j}\dot{q}_{j})q_{i} - (\dot{q}^{j} q_{j})\dot{q}_{i} - \frac{k}{r(b_{0}+b_{1}t)}q_{i}$.

\begin{longtable}{|c|c|l|}
\hline
$\nu$ & $\omega(t)$ & LFIs and QFIs \\ \hline
\multirow{3}{*}{$\forall$ $\nu$} & $\forall$ $\omega$ & $L_{i} = q_{i+1}\dot{q}_{i+2} - q_{i+2}\dot{q}_{i+1}$, $L_{ij}= q_{i}\dot{q}_{j} - q_{j}\dot{q}_{i} =\varepsilon_{ijk} L^{k}$ \\
& $k$ & $H_{\nu}= \frac{1}{2}\dot{q}^{i}\dot{q}_{i} - \frac{k}{r^{\nu}}$ \\
& $\omega_{\nu}= k\left(b_{0} + b_{1}t + b_{2}t^{2} \right)^{\frac{\nu-2}{2}}$ & $J_{\nu}=  (b_{0} + b_{1}t + b_{2}t^{2}) \left( \frac{\dot{q}^{i}\dot{q}_{i}}{2} - \frac{\omega_{\nu}}{r^{\nu}} \right) -\frac{b_{1} + 2b_{2}t}{2} q^{i}\dot{q}_{i} +\frac{b_{2} r^{2}}{2}$ \\ \hline
\multirow{10}{*}{$-2$} & $k$ & $B_{ij} = \dot{q}_{i} \dot{q}_{j} - 2k q_{i}q_{j}$ \\
& $k>0$ & $I_{3a\pm}= e^{\pm \sqrt{2k} t}(\dot{q}_{a} \mp \sqrt{2k} q_{a})$ \\
& $k<0$ & $I_{3a\pm}= e^{\pm i \sqrt{-2k} t}(\dot{q}_{a} \mp i \sqrt{-2k} q_{a})$ \\
& $\frac{k}{(b_{0} +b_{1}t +b_{2}t^{2})^{2}}$ & $I_{ij}= (b_{0}+b_{1}t+b_{2}t^{2}) \left( \dot{q}_{i}\dot{q}_{j} -2\omega q_{i}q_{j} \right) -(b_{1} +2b_{2}t)q_{(i}\dot{q}_{j)} +b_{2}q_{i}q_{j}$ \\ \cline{2-3}
& $\frac{\ddot{f}}{4f(t)} - \frac{1}{8} \left(\frac{\dot{f}}{f}\right)^{2} -\frac{c_{0}}{4f^{2}}$ & \makecell[l]{$L_{ij}= \left( \frac{2}{c_{0}} \right)^{1/2} \left( I_{41i}I_{42j} - I_{41j}I_{42i} \right)$, \\ $\Lambda_{ij}= f(t) \left( \dot{q}_{i}\dot{q}_{j} -2\omega q_{i} q_{j} \right) -\dot{f} q_{(i}\dot{q}_{j)} +\frac{\ddot{f}}{2} q_{i}q_{j} = I_{41i}I_{41j} + I_{42i}I_{42j}$, \\ $I_{41i} =\left( \frac{c_{0}}{2} \right)^{1/2} f^{-1/2}q_{i} \sin\theta + \left( f^{1/2} \dot{q}_{i} - \frac{\dot{f}}{2} f^{-1/2}q_{i} \right) \cos\theta$, \\ $I_{42i}= -\left( \frac{c_{0}}{2} \right)^{1/2} f^{-1/2}q_{i} \cos\theta + \left( f^{1/2} \dot{q}_{i} - \frac{\dot{f}}{2} f^{-1/2}q_{i} \right) \sin\theta$ \\ where $\theta=  \left( \frac{c_{0}}{2} \right)^{1/2} \int f^{-1}dt$} \\ \cline{2-3}
& $\frac{\ddot{g}}{2g(t)}$ & $I_{4i}= g(t)\dot{q}_{i} -\dot{g}q_{i}$ \\ \hline
\multirow{5}{*}{$1$} & $k$ & $R_{i}= (\dot{q}^{j} \dot{q}_{j}) q_{i} - (\dot{q}^{j}q_{j})\dot{q}_{i}- \frac{k}{r}q_{i}$ \\ \cline{2-3}
& $\frac{k}{b_{0}+b_{1}t}$ & \makecell[l]{$E_{2}= (b_{0}+b_{1}t)^{2} \left[ \frac{\dot{q}^{i} \dot{q}_{i}}{2} -\frac{k}{r(b_{0}+b_{1}t)} \right] -b_{1}(b_{0}+b_{1}t) q^{i}\dot{q}_{i} + \frac{b_{1}^{2}r^{2}}{2}$, \\ $A_{i}= (b_{0}+b_{1}t) \tilde{R}_{i} + b_{1}\left( q_{i+2}L_{i+1} - q_{i+1}L_{i+2} \right)$ \\ where $\tilde{R}_{i}= (\dot{q}^{j} \dot{q}_{j})q_{i} - (\dot{q}^{j}q_{j})\dot{q}_{i} - \frac{k}{r(b_{0}+b_{1}t)} q_{i}$} \\ \cline{2-3}
& $\frac{k}{(b_{0}+b_{1}t+b_{2}t^{2})^{1/2}}$ & $E_{3} = (b_{0}+b_{1}t+b_{2}t^{2}) \left[ \frac{\dot{q}^{i} \dot{q}_{i}}{2} -\frac{k}{r ( b_{0}+b_{1}t +b_{2}t^{2} )^{1/2}} \right] - \frac{b_{1}+2b_{2}t}{2} q^{i}\dot{q}_{i} + \frac{b_{2}r^{2}}{2}$ \\
\hline
$2$ & $k$ & $I_{1}= - H_{2}t^{2} + t(\dot{q}^{i}q_{i}) - \frac{r^{2}}{2}$, $I_{2}= - H_{2}t + \frac{1}{2} (\dot{q}^{i}q_{i})$ \\ \hline
\caption{\label{T2} The LFIs/QFIs of the time-dependent generalized Kepler potential $V=-\frac{\omega(t)}{r^{\nu}}$.}
\end{longtable}

\section{Integrating the equations}

\label{sec.applications.GK}

In this section we use the independent LFIs $I_{41i}, I_{42i}$ to integrate the equations of the special class of 3d time-dependent oscillators ($\nu= -2$) defined in section \ref{sec.discussion.GK} with $\omega(t)$ given by (\ref{eq.disosc3}). We also use the FIs $L_{i}$, $E_{2}$, $A_{i}$ to integrate the time-dependent Kepler potential ($\nu=1$) with $\omega(t)= \frac{k}{b_{0}+b_{1}t}$ where $kb_{1}\neq0$ (see subsection \ref{sec.omega.2}).

\subsection{The 3d time-dependent oscillator with $\omega(t)$ given by (\ref{eq.disosc3})}

\label{sec.exa.oscillator}

Using the LFIs (\ref{eq.disosc4b}) and (\ref{eq.disosc4c}) we find
\begin{equation}
q_{i}(t) = \left( \frac{2}{c_{0}} \right)^{1/2} f^{1/2} \Big( I_{41i}\sin\theta - I_{42i}\cos\theta \Big) \label{eq.exaGK.1}
\end{equation}
where $I_{41i}, I_{42i}$, $i=1,2,3$, are arbitrary constants (real or imaginary) and $\theta(t)=  \left( \frac{c_{0}}{2} \right)^{1/2} \int f^{-1}dt$.

The solution (\ref{eq.exaGK.1}) coincides with the solution (52) in \cite{Prince 1980}.

In the case of the 1d time-dependent oscillator, if we set $2\omega(t)=-\psi^{2}(t)$, $c_{0}=2$ and $f(t)=\rho^{2}(t)$, equation (\ref{eq.TGKep.1a}) and the defining relation (\ref{eq.disosc3}) for $\omega(t)$ become
\begin{eqnarray}
\ddot{x} &=& -\psi^{2}x \label{eq.ermexa1} \\
\ddot{\rho} &=& -\psi^{2}\rho + \rho^{-3}. \label{eq.ermexa2}
\end{eqnarray}
The LFIs (\ref{eq.disosc9a}) and (\ref{eq.disosc9b}) become
\begin{eqnarray}
I_{41} &=& \rho^{-1}x \sin\theta + \left( \rho \dot{x} - x\dot{\rho} \right) \cos\theta \label{eq.ermexa3} \\
I_{42} &=& -\rho^{-1}x \cos\theta + \left( \rho \dot{x} - x\dot{\rho} \right) \sin\theta. \label{eq.ermexa4}
\end{eqnarray}
The general solution (\ref{eq.exaGK.1}) is
\begin{equation}
x(t) = \rho(t) \Big( I_{41}\sin\theta - I_{42}\cos\theta \Big) \label{eq.ermexa5}
\end{equation}
where $\dot{\theta}= \rho^{-2}$ and $\rho(t)$ is a given non-zero function which defines $\psi(t)$ through (\ref{eq.ermexa2}). This is the 1d solution (9) in \cite{Prince 1980}.

\subsection{The solution of the time-dependent Kepler potential with $\omega_{2K}(t)= \frac{k}{b_{0}+b_{1}t}$ where $kb_{1}\neq0$}

In subsection \ref{sec.omega.2} it is shown that this system admits the following FIs:
\[
L_{1}=y\dot{z} -z\dot{y}, \enskip L_{2}= z\dot{x} -x\dot{z}, \enskip L_{3}= x\dot{y} -y\dot{x}
\]
\[
E_{2}= (b_{0}+b_{1}t)^{2} \left[ \frac{\dot{q}^{i} \dot{q}_{i}}{2} -\frac{k}{r(b_{0}+b_{1}t)} \right] -b_{1}(b_{0}+b_{1}t) q^{i}\dot{q}_{i} + \frac{b_{1}^{2}r^{2}}{2}
\]
\[
A_{i}= (b_{0}+b_{1}t) \tilde{R}_{i} + b_{1}\left( q_{i+2}L_{i+1} - q_{i+1}L_{i+2} \right)
\]
where $\tilde{R}_{i}= (\dot{q}^{j}\dot{q}_{j})q_{i} - (\dot{q}^{j}q_{j})\dot{q}_{i} - \frac{k}{r(b_{0}+b_{1}t)} q_{i}$. The components of the generalized Runge-Lenz vector are written
\begin{eqnarray*}
A_{1}&=& (b_{0}+b_{1}t)(\dot{y}L_{3} -\dot{z}L_{2}) + b_{1}\left( zL_{2} - yL_{3} \right) - \frac{k}{r}x \\
A_{2}&=& (b_{0}+b_{1}t)(\dot{z}L_{1} -\dot{x}L_{3}) + b_{1}\left( xL_{3} - zL_{1} \right) - \frac{k}{r}y \\
A_{3}&=& (b_{0}+b_{1}t)(\dot{x}L_{2} -\dot{y}L_{1}) + b_{1}\left( yL_{1} - xL_{2} \right) -\frac{k}{r}z.
\end{eqnarray*}

Since the angular momentum is a FI the motion is on a plane. We choose without loss of generality the plane $z=0$ and on that the polar coordinates $x=r\cos\theta$, $y=r\sin\theta$. Then
\[
L_{1}=L_{2}=0, \enskip L_{3}=r^{2}\dot{\theta}, \enskip E_{2}= (b_{0}+b_{1}t)^{2} \left[ \frac{\dot{r}^{2} +r^{2} \dot{\theta}^{2}}{2} -\frac{k}{r(b_{0}+b_{1}t)} \right] -b_{1}(b_{0}+b_{1}t) r\dot{r} + \frac{b_{1}^{2}r^{2}}{2}
\]
\[
A_{1}= L_{3} \Big[ (b_{0}+b_{1}t)\dot{r} -b_{1}r \Big] \sin\theta + \Big[ (b_{0}+b_{1}t)L_{3}r\dot{\theta} - k \Big] \cos\theta
\]
\[
A_{2}= -L_{3} \Big[ (b_{0}+b_{1}t)\dot{r} -b_{1}r \Big] \cos\theta + \Big[ (b_{0}+b_{1}t) L_{3} r\dot{\theta} -k \Big]\sin\theta, \enskip  A_{3}=0.
\]

Using the relation $\dot{\theta}=\frac{L_{3}}{r^{2}}$ to replace $\dot{\theta}$, the above relations are written
\begin{eqnarray}
E_{2}&=& (b_{0}+b_{1}t)^{2} \left[ \frac{\dot{r}^{2}}{2} + \frac{L_{3}^{2}}{2r^{2}} -\frac{k}{r(b_{0}+b_{1}t)} \right] -b_{1}(b_{0}+b_{1}t) r\dot{r} + \frac{b_{1}^{2}r^{2}}{2} \label{eq.exaGK.2a} \\
A_{1}&=& L_{3} \Big[ (b_{0}+b_{1}t)\dot{r} -b_{1}r \Big] \sin\theta + \Big[ (b_{0}+b_{1}t)\frac{L_{3}^{2}}{r} - k \Big] \cos\theta \label{eq.exaGK.2b} \\
A_{2}&=& -L_{3} \Big[ (b_{0}+b_{1}t)\dot{r} -b_{1}r \Big] \cos\theta + \Big[ (b_{0}+b_{1}t) \frac{L_{3}^{2}}{r} -k \Big] \sin\theta. \label{eq.exaGK.2c}
\end{eqnarray}

By multiplying equation (\ref{eq.exaGK.2b}) with $\cos\theta$ and (\ref{eq.exaGK.2c}) with $\sin\theta$ we find that
\begin{equation}
\frac{1}{r} = \frac{k}{L_{3}^{2}(b_{0}+b_{1}t)} \left( 1 + k_{1}\cos\theta + k_{2} \sin\theta \right) \implies r= \frac{L_{3}^{2}(b_{0}+b_{1}t)}{k\left( 1 + k_{1}\cos\theta + k_{2} \sin\theta \right)} \label{eq.exaGK.3}
\end{equation}
where $k_{1}\equiv \frac{A_{1}}{k}$ and $k_{2}\equiv \frac{A_{2}}{k}$.

Applying the transformation $k_{1}= \alpha \cos\beta$ and $k_{2}= \alpha \sin\beta$, equation (\ref{eq.exaGK.3}) is written (see also section 5 in \cite{Katzin 1982})
\begin{equation}
\frac{1}{r} = \frac{\omega_{2K}}{L_{3}^{2}} \Big[ 1 + \alpha \cos \left( \theta -\beta \right) \Big] \implies r = \frac{L_{3}^{2}\omega_{2K}^{-1}}{1 + \alpha \cos \left(\theta - \beta\right)} \label{eq.exaGK.4}
\end{equation}
which for $\omega_{2K}(t)=const$ (standard Kepler problem) reduces to the analytical equation of a conic section in polar coordinates. In that case $\alpha$ is the eccentricity.

It is also worthwhile to mention that the relation (\ref{eq.TKFI4}) becomes
\[
2E_{2}L_{3}^{2} + k^{2}= \alpha^{2}k^{2} \implies 2E_{2}L_{3}^{2} = k^{2}(\alpha^{2}-1).
\]

Moreover, equation (\ref{eq.exaGK.2a}) gives
\[
\left[ \frac{d}{dt} \left( \frac{r}{b_{0}+b_{1}t} \right) \right]^{2} = -2(b_{0}+b_{1}t)^{-2} \left[ \frac{L_{3}^{2}}{2r^{2}} -\frac{k}{r(b_{0}+b_{1}t)} -\frac{E_{2}}{(b_{0}+b_{1}t)^{2}} \right].
\]

Finally, in the polar plane the equations of motion (\ref{eq.TGKep.1a}) for $\nu=1$ become
\begin{eqnarray}
\ddot{r} - r\dot{\theta}^{2} +\frac{\omega_{2K}}{r^{2}} &=& 0 \label{eq.exaGK.5a} \\
r\ddot{\theta} + 2\dot{r}\dot{\theta}&=&0. \label{eq.exaGK.5b}
\end{eqnarray}
Equation (\ref{eq.exaGK.5b}) implies the FI of the angular momentum $L_{3} =r^{2}\dot{\theta}$. It can be easily checked that the solution (\ref{eq.exaGK.3}) satisfies equation (\ref{eq.exaGK.5a}) by replacing $\ddot{\theta}$ from (\ref{eq.exaGK.5b}) and $\dot{\theta}$ with $\frac{L_{3}}{r^{2}}$. The solution (\ref{eq.exaGK.3}) into the FI $L_{3}$ gives
\begin{equation}
\int \frac{k^{2}dt}{L_{3}^{3}(b_{0}+b_{1}t)^{2}}= \int \frac{d\theta}{\left( 1 + k_{1}\cos\theta + k_{2} \sin\theta \right)^{2}} \implies \frac{k}{L_{3}^{2} (b_{0}+b_{1}t)} = - \frac{b_{1} L_{3}}{k}\int \frac{d\theta}{\left( 1 + k_{1}\cos\theta + k_{2} \sin\theta \right)^{2}}. \label{eq.exaGK.6}
\end{equation}
Substituting (\ref{eq.exaGK.6}) in (\ref{eq.exaGK.3}) we obtain
\begin{equation}
\frac{1}{r} = - \frac{b_{1}L_{3}}{k} \left( 1 + k_{1}\cos\theta + k_{2} \sin\theta \right) \int \frac{d\theta}{\left( 1 + k_{1}\cos\theta + k_{2} \sin\theta \right)^{2}} \label{eq.exaGK.7}
\end{equation}
which coincides with eq. (5.17) in \cite{Katzin 1982}.

\section{A class of 1d non-linear time-dependent equations}

\label{sec.nonlin}

In this section we use the well-known result \cite{LeoTsampAndro 2017} that the non-linear dynamical system
\begin{equation}
\ddot{q}^{a}= -\Gamma^{a}_{bc}\dot{q}^{b}\dot{q}^{c} -\omega(t)Q^{a}(q) +\phi(t)\dot{q}^{a} \label{eq.damp0a}
\end{equation}
is equivalent to the linear dynamical system (without damping term)
\begin{equation}
\frac{d^{2}q^{a}}{ds^{2}}= -\Gamma^{a}_{bc}\frac{dq^{b}}{ds} \frac{dq^{c}}{ds} -\bar{\omega}(s)Q^{a}(q) \label{eq.damp0b}
\end{equation}
where $\phi(t)$ is an arbitrary function such that
\begin{equation}
s(t)= \int e^{\int\phi(t)dt} dt, \enskip \bar{\omega}(s)= \omega(t(s)) \left(\frac{dt}{ds}\right)^{2} \iff \omega(t)= \bar{\omega}(s(t)) e^{2\int\phi(t)dt}. \label{eq.damp0c}
\end{equation}

We apply this result to the following problem:

\emph{Consider the second order differential equation}
\begin{equation}
\ddot{x}=-\omega(t)x^{\mu }+\phi(t)\dot{x}  \label{eq.nonl1}
\end{equation}%
\emph{where the constant }$\mu\neq-1$\emph{ and determine the relation
between the functions }$\omega(t), \phi (t)$\emph{\ for which the equation
admits a QFI, therefore it is integrable.}

This problem has been considered previously in \cite{Da Silva 1974}, \cite{Sarlet 1980} (see eq. (28a) in \cite{Da Silva 1974} and eq. (17) in \cite{Sarlet 1980}) and has been answered partially using different methods. In \cite{Da Silva 1974} the author used the Hamiltonian formalism where one looks for a canonical transformation to bring the Hamiltonian in a time-separable form. In \cite{Sarlet 1980} the author used a direct method for constructing FIs by multiplying the equation with an integrating factor. In \cite{Sarlet 1980} it is shown that both methods are equivalent and that the results of \cite{Sarlet 1980} generalize those of \cite{Da Silva 1974}. In the following we shall generalize the results of \cite{Sarlet 1980}; in addition we discuss a number of applications.

Equation (\ref{eq.nonl1}) is equivalent to the equation
\begin{equation}
\frac{d^{2}x}{ds^{2}}= -\bar{\omega}(s)x^{\mu}, \enskip \mu\neq-1 \label{eq.nonl2}
\end{equation}
where the function $\bar{\omega}(s)$ is given by (\ref{eq.damp0c}).

Replacing with $Q^{1}=x^{\mu}$ in the system of equations (\ref{eq.TKN1}) - (\ref{eq.TKN6}) we find that\footnote{
In 1d Euclidean space the KT condition (\ref{eq.TKN1}) $K_{(ab;c)}=0$ becomes $K_{11,1}=0$ $\implies K_{11}=K_{11}(s)$, that is, it is an arbitrary function of $s$.
} $K_{11}= K_{11}(s)$ and the following conditions
\begin{eqnarray}
K_{1}(s,x) &=& -\frac{dK_{11}}{ds}x + b_{1}(s) \label{eq.nonl3a} \\
K(s,x) &=& 2\bar{\omega} K_{11} \frac{x^{\mu+1}}{\mu+1} + \frac{d^{2}K_{11}}{ds^{2}} \frac{x^{2}}{2} -\frac{db_{1}}{ds}x +b_{2}(s) \label{eq.nonl3b} \\
0 &=& \left( \frac{2\frac{d\bar{\omega}}{ds} K_{11}}{\mu+1} +\frac{2\bar{\omega} \frac{dK_{11}}{ds}}{\mu+1} +\bar{\omega}\frac{dK_{11}}{ds} \right) x^{\mu+1} -\bar{\omega} b_{1} x^{\mu} + \frac{d^{3}K_{11}}{ds^{3}}\frac{x^{2}}{2} -\frac{d^{2}b_{1}}{ds^{2}}x + \frac{db_{2}}{ds} \label{eq.nonl3c}
\end{eqnarray}
where $b_{1}(s), b_{2}(s)$ are arbitrary functions. Then the general QFI (\ref{FI.5}) becomes
\begin{equation}
I= K_{11}(s) \left( \frac{dx}{ds} \right)^{2} +K_{1}(s,x)\frac{dx}{ds} + K(s,x). \label{eq.nonl4}
\end{equation}

We consider the solution of the system (\ref{eq.nonl3a}) - (\ref{eq.nonl3c}) for various values of $\mu$.

As will be shown for $\mu \neq -1$ results a family of `frequencies' $\bar{\omega}(s)$ parameterized with constants. However, for the specific values $\mu =0,1,2$ there results a family of `frequencies' $\bar{\omega}(s)$ parameterized with functions.
\bigskip

1) Case $\mu=0$.

We find the QFI
\begin{equation}
I= K_{11} \left( \frac{dx}{ds} \right)^{2} - \frac{dK_{11}}{ds} x\frac{dx}{ds} +b_{1}(s)\frac{dx}{ds} + c_{3}x^{2} +2\bar{\omega}(s)K_{11}x -\frac{db_{1}}{ds}x + \int b_{1}(s) \bar{\omega}(s) ds \label{eq.nonl4.1}
\end{equation}
where $K_{11}= c_{1} +c_{2}s + c_{3}s^{2}$, $c_{1}, c_{2}, c_{3}$ are arbitrary constants and the functions $b_{1}(s), \bar{\omega}(s)$ satisfy the condition
\begin{equation}
\frac{d^{2}b_{1}}{ds^{2}}= 2\frac{d\bar{\omega}}{ds}K_{11} +3\bar{\omega}\frac{dK_{11}}{ds}. \label{eq.nonl4.2}
\end{equation}

Using the transformation (\ref{eq.damp0c}) equations (\ref{eq.nonl4.1}), (\ref{eq.nonl4.2}) become
\begin{eqnarray}
I&=& \left[ c_{1} +c_{2}\int e^{\int\phi(t)dt} dt +c_{3}\left(\int e^{\int\phi(t)dt} dt\right)^{2} \right] e^{-2\int \phi(t)dt} \dot{x}^{2}  -\left[ c_{2} +2c_{3}\int e^{\int\phi(t)dt} dt \right] e^{-\int \phi(t)dt} x\dot{x} + \notag\\
&& +b_{1}(s(t)) e^{-\int \phi(t)dt} \dot{x} + c_{3}x^{2} + 2\omega(t) \left[ c_{1} +c_{2}\int e^{\int\phi(t)dt} dt +c_{3}\left(\int e^{\int\phi(t)dt} dt\right)^{2} \right] e^{-2\int \phi(t)dt} x - \notag \\
&& -\dot{b}_{1} e^{-\int \phi(t)dt}x + \int b_{1}(s(t)) \omega(t) e^{-\int \phi(t)dt} dt \label{eq.nonl4.2.1}
\end{eqnarray}
and
\begin{eqnarray}
\ddot{b}_{1} -\phi\dot{b}_{1} &=& 2e^{-\int \phi(t)dt} \left( \dot{\omega} -2\phi \omega \right) \left[ c_{1} +c_{2}\int e^{\int\phi(t)dt} dt +c_{3}\left(\int e^{\int\phi(t)dt} dt\right)^{2} \right] + \notag \\
&& + 3\omega \left[ c_{2} +2c_{3}\int e^{\int\phi(t)dt} dt \right]. \label{eq.nonl4.2.2}
\end{eqnarray}

2) Case $\mu=1$.

We derive again the results of the time-dependent oscillator (see Table \ref{T2} for $\nu=-2$) in one dimension. Using the transformation (\ref{eq.damp0c}) we deduce that the original equation
\begin{equation}
\ddot{x}= -\omega(t)x +\phi(t)\dot{x} \label{eq.nonl4.2.5}
\end{equation}
for the frequency
\begin{equation}
\omega(t)= -\rho^{-1}\ddot{\rho} +\phi (\ln \rho)^{\cdot} +\rho^{-4} e^{2\int\phi(t)dt} \label{eq.nonl4.2.6}
\end{equation}
admits the general solution
\begin{equation}
x(t)= \rho(t) \left( A \sin\theta + B\cos\theta \right) \label{eq.nonl4.2.7}
\end{equation}
where $\rho(t)\equiv \rho(s(t))$ and $\theta(s(t))= \int \rho^{-2}(t) e^{\int\phi(t)dt}dt$.

3) Case $\mu=2$.

We find the function $\bar{\omega}= K_{11}^{-5/2}$ and the QFI
\begin{equation}
I= K_{11}(s) \left( \frac{dx}{ds} \right)^{2} -\frac{dK_{11}}{ds} x\frac{dx}{ds} +(c_{4}+c_{5}s) \frac{dx}{ds} + \frac{2}{3}K_{11}^{-3/2} x^{3} + \frac{d^{2}K_{11}}{ds^{2}} \frac{x^{2}}{2} -c_{5}x \label{eq.nonl4.3}
\end{equation}
where $c_{4}, c_{5}$ are arbitrary constants and the function $K_{11}(s)$ is given by
\begin{equation}
\frac{d^{3}K_{11}}{ds^{3}}= 2(c_{4}+c_{5}s)K_{11}^{-5/2}. \label{eq.nonl4.4}
\end{equation}

Using the transformation (\ref{eq.damp0c}) the above results become
\begin{equation}
\omega(t)= K_{11}^{-5/2} e^{2\int\phi(t)dt} \label{eq.nonl4.4.1}
\end{equation}
\begin{eqnarray}
I&=& K_{11} e^{-2\int \phi(t)dt} \dot{x}^{2} -\dot{K}_{11} e^{-2\int \phi(t)dt} x\dot{x} +\left[ c_{4} +c_{5}\int e^{\int\phi(t)dt} dt \right] e^{-\int \phi(t)dt}\dot{x} + \frac{2}{3}K_{11}^{-3/2} x^{3} + \notag \\
&&+ \left( \ddot{K}_{11} -\phi\dot{K}_{11} \right)e^{-2\int \phi(t)dt} \frac{x^{2}}{2} -c_{5}x \label{eq.nonl4.4.2}
\end{eqnarray}
and
\begin{equation}
\dddot{K}_{11} -3\phi \ddot{K}_{11} -\dot{\phi}\dot{K}_{11} +2\phi^{2}\dot{K}_{11}= 2\left[ c_{4} +c_{5}\int e^{\int\phi(t)dt} dt \right] e^{3\int\phi(t)dt} K_{11}^{-5/2} \label{eq.nonl4.4.3}
\end{equation}
where the function $K_{11}=K_{11}(s(t))$.

We note that for $\mu=2$ equation (\ref{eq.nonl1}), or to be more specific its equivalent (\ref{eq.nonl2}), arises in the solution of Einstein field equations when the gravitational field is spherically symmetric and the matter source is a shear-free perfect fluid (see e.g. \cite{StephaniB}, \cite{Stephani 1983}, \cite{Srivastana 1987}, \cite{Leach 1992}, \cite{LeachMaartens 1992}, \cite{Maharaj 1996}).

4) Case $\mu \neq -1$.

In this case $b_{1}=b_{2}=0$, $K_{11}= c_{1} +c_{2}s +c_{3}s^{2}$ and $\bar{\omega}(s)= (c_{1} +c_{2}s +c_{3}s^{2})^{-\frac{\mu+3}{2}}$ where $c_{1}, c_{2}, c_{3}$ are arbitrary constants.

The QFI (\ref{eq.nonl4}) becomes
\begin{equation}
I= (c_{1} +c_{2}s +c_{3}s^{2})\left( \frac{dx}{ds} \right)^{2}  -(c_{2} +2c_{3}s)x\frac{dx}{ds} +\frac{2}{\mu+1} (c_{1} +c_{2}s +c_{3}s^{2})^{-\frac{\mu +1}{2}} x^{\mu+1} + c_{3}x^{2} \label{eq.nonl5}
\end{equation}
and the function
\begin{equation}
\bar{\omega}(s)= (c_{1} +c_{2}s +c_{3}s^{2})^{-\frac{\mu +3}{2}}. \label{eq.nonl6}
\end{equation}
It can be checked that (\ref{eq.nonl5}), (\ref{eq.nonl6}) for $\mu=0, 1, 2$ give results compatible with the ones we found for these values of $\mu$.

Using the transformation (\ref{eq.damp0c}) we deduce that the original system (\ref{eq.nonl1}) is integrable iff the functions $\omega(t), \phi(t)$ are related as follows
\begin{equation}
\omega(t)= \left[ c_{1} +c_{2}\int e^{\int\phi(t)dt} dt +c_{3}\left(\int e^{\int\phi(t)dt} dt\right)^{2} \right]^{-\frac{\mu+3}{2}} e^{2\int \phi(t)dt}. \label{eq.nonl6.1}
\end{equation}
In this case the associated QFI (\ref{eq.nonl5}) is
\begin{eqnarray}
I&=& \left[ c_{1} +c_{2}\int e^{\int\phi(t)dt} dt +c_{3}\left(\int e^{\int\phi(t)dt} dt\right)^{2} \right] e^{-2\int \phi(t)dt} \dot{x}^{2} -\left[ c_{2} +2c_{3}\int e^{\int\phi(t)dt} dt \right] e^{-\int \phi(t)dt} x\dot{x} + \notag \\
&& +\frac{2}{\mu+1} \left[ c_{1} +c_{2}\int e^{\int\phi(t)dt} dt +c_{3}\left(\int e^{\int\phi(t)dt} dt\right)^{2} \right]^{-\frac{\mu +1}{2}} x^{\mu+1} + c_{3}x^{2}. \label{eq.nonl6.2}
\end{eqnarray}

These expressions generalize the ones given in \cite{Sarlet 1980}. Indeed if we introduce the notation $\omega(t)\equiv \alpha(t)$, $\phi(t)\equiv -\beta(t)$, then equations (\ref{eq.nonl6.1}), (\ref{eq.nonl6.2}) for $c_{3}=0$ become eqs. (25), (26) of \cite{Sarlet 1980}.

\subsection{The generalized Lane-Emden equation}

\label{sec.emden}

Consider the 1d generalized Lane-Emden equation (see eq. (6) in \cite{Muatje 2011})
\begin{equation}
\ddot{x}= -\omega(t)x^{\mu} -\frac{k}{t}\dot{x} \label{eq.emd1}
\end{equation}
where $k$ is an arbitrary constant. This equation is well-known in the literature because of its many applications in astrophysical problems (see Refs. in \cite{Muatje 2011}). In general, to find explicit analytic solutions of equation (\ref{eq.emd1}) is a major task. For example, such solutions have been found only for the special values $\mu=0,1,5$, in the case that the function $\omega(t)=1$ and the constant $k=2$. New exact solutions, or at least the Liouville integrability, of equation (\ref{eq.emd1}) are guaranteed, if we find a way to determine its FIs. We see that equation (\ref{eq.emd1}) is a subcase of the original equation (\ref{eq.nonl1}) for $\phi(t)= -\frac{k}{t}$, therefore we can apply the results found earlier in section \ref{sec.nonlin}.

In what follows we discuss only the fourth case where $\mu\neq-1$ in order to compare our results with those found in Table 1 of \cite{Muatje 2011}. In particular, for $\phi(t)= -\frac{k}{t}$ the function (\ref{eq.nonl6.1}) and the associated QFI (\ref{eq.nonl6.2}) become
\begin{equation}
\omega(t)= t^{-2k} \left( c_{1} +c_{2}M +c_{3}M^{2} \right)^{-\frac{\mu+3}{2}} \label{eq.emd2}
\end{equation}
and
\begin{equation}
I= t^{2k} \left( c_{1} +c_{2}M +c_{3}M^{2} \right) \dot{x}^{2} -t^{k} \left( c_{2} +2c_{3}M \right) x\dot{x} +\frac{2}{\mu+1} \left( c_{1} +c_{2}M +c_{3}M^{2} \right)^{-\frac{\mu +1}{2}} x^{\mu+1} + c_{3}x^{2} \label{eq.emd3}
\end{equation}
where the function $M(t)= \int t^{-k}dt$.

Concerning the form of the function $M(t)$ there are two cases to be considered: a) $k=1$; and b) $k\neq1$.
\bigskip

a) Case $k=1$.

We have $M=\ln t$ and equations (\ref{eq.emd2}), (\ref{eq.emd3}) give
\begin{equation}
\omega(t)= t^{-2} \left[ c_{1} +c_{2}\ln t +c_{3}(\ln t)^{2} \right]^{-\frac{\mu+3}{2}} \label{eq.emd4.1}
\end{equation}
and
\begin{eqnarray}
I&=& t^{2} \left[ c_{1} +c_{2}\ln t +c_{3}(\ln t)^{2} \right] \dot{x}^{2} -t \left( c_{2} +2c_{3}\ln t \right) x\dot{x} +\notag \\
&& +\frac{2}{\mu+1} \left[ c_{1} +c_{2}\ln t +c_{3}(\ln t)^{2} \right]^{-\frac{\mu +1}{2}} x^{\mu+1} + c_{3}x^{2}. \label{eq.emd4.2}
\end{eqnarray}

We consider the following subcases:

-$c_{2}=c_{3}=0$, $c_{1}\neq0$.

Equations (\ref{eq.emd4.1}), (\ref{eq.emd4.2}) give the function $\omega(t)= At^{-2}$ and the QFI (divide $I$ with $2c_{1}$)
\[
I= \frac{t^{2}}{2}\dot{x}^{2} + \frac{A}{\mu+1}x^{\mu+1}
\]
where the constant $A=c_{1}^{-\frac{\mu+3}{2}}$. This is the Case 5 in Table 1 of \cite{Muatje 2011}.

- $c_{1}=c_{3}=0$, $c_{2}\neq0$.

Equations (\ref{eq.emd4.1}), (\ref{eq.emd4.2}) give the function $\omega(t)= At^{-2}(\ln t)^{-\frac{\mu+3}{2}}$ and the QFI (divide $I$ with $2c_{2}$)
\[
I= \frac{1}{2}t^{2}(\ln t)\dot{x}^{2} -\frac{t}{2} x\dot{x} +\frac{A}{\mu+1} (\ln t)^{-\frac{\mu+1}{2}} x^{\mu+1}
\]
where the constant $A= c_{2}^{-\frac{\mu+3}{2}}$. This is the Case 6 in Table 1 of \cite{Muatje 2011}.

- $c_{1}=c_{2}=0$, $c_{3}\neq0$.

Equations (\ref{eq.emd5.1}), (\ref{eq.emd5.2}) give the function $\omega(t)= At^{-2}(\ln t)^{-\mu-3}$ and the QFI (divide $I$ with $2c_{3}$)
\[
I= \frac{1}{2} (t\ln t)^{2} \dot{x}^{2} -t(\ln t)x\dot{x} +\frac{A}{\mu+1} (\ln t)^{-\mu-1}x^{\mu+1} +\frac{x^{2}}{2}
\]
where the constant $A= c_{3}^{-\frac{\mu+3}{2}}$. This is the Case 7 in Table 1 of \cite{Muatje 2011}.

\bigskip

b) Case $k\neq1$.

We have $M= \frac{t^{1-k}}{1-k}$ and equations (\ref{eq.emd2}), (\ref{eq.emd3}) give
\begin{equation}
\omega(t)= t^{-2k} \left[ c_{1} +\frac{c_{2}}{1-k}t^{1-k} +\frac{c_{3}}{(1-k)^{2}}t^{2(1-k)} \right]^{-\frac{\mu+3}{2}} \label{eq.emd5.1}
\end{equation}
and
\begin{eqnarray}
I&=& t^{2k} \left[ c_{1} +\frac{c_{2}}{1-k}t^{1-k} +\frac{c_{3}}{(1-k)^{2}}t^{2(1-k)} \right] \dot{x}^{2} -t^{k} \left( c_{2} +\frac{2c_{3}}{1-k}t^{1-k} \right) x\dot{x} +\notag \\
&& +\frac{2}{\mu+1} \left[ c_{1} +\frac{c_{2}}{1-k}t^{1-k} +\frac{c_{3}}{(1-k)^{2}}t^{2(1-k)} \right]^{-\frac{\mu +1}{2}} x^{\mu+1} + c_{3}x^{2}. \label{eq.emd5.2}
\end{eqnarray}

We consider the following subcases:

- $c_{2}=c_{3}=0$, $c_{1}\neq0$.

Equations (\ref{eq.emd5.1}), (\ref{eq.emd5.2}) give the function $\omega(t)= At^{-2k}$ and the QFI (divide $I$ with $2c_{1}$)
\[
I= \frac{t^{2k}}{2}\dot{x}^{2} + \frac{A}{\mu+1}x^{\mu+1}
\]
where the constant $A=c_{1}^{-\frac{\mu+3}{2}}$. This is the Case 2 in Table 1 of \cite{Muatje 2011}.

- $c_{1}=c_{3}=0$, $c_{2}\neq0$.

Equations (\ref{eq.emd5.1}), (\ref{eq.emd5.2}) give the function $\omega(t)= At^{\frac{1}{2}(k\mu -k -\mu -3)}$ and the QFI (multiply $I$ with $\frac{1-k}{c_{2}}$)
\[
I= t^{k+1}\dot{x}^{2} +(k-1)t^{k}x\dot{x} +\frac{2A}{\mu+1} t^{\frac{1}{2}(\mu+1)(k-1)} x^{\mu+1}
\]
where the constant $A= \left( \frac{c_{2}}{1-k} \right)^{-\frac{\mu+3}{2}}$. This is the Case 3 in Table 1 of \cite{Muatje 2011}.

We note also that for $k=\frac{\mu+3}{\mu-1}$ where $\mu\neq1$  the function $\omega(t)=A=const$. This reproduces the first subcase of Case 1 in Table 1 of \cite{Muatje 2011} which is the Case 5.1 of \cite{Khalique 2008}.

- $c_{1}=c_{2}=0$, $c_{3}\neq0$.

Equations (\ref{eq.emd5.1}), (\ref{eq.emd5.2}) give the function $\omega(t)= At^{k\mu +k -\mu -3}$ and the QFI (multiply $I$ with $\frac{(1-k)^{2}}{2c_{3}}$)
\[
I= \frac{t^{2}}{2}\dot{x}^{2} +(k-1)tx\dot{x} +\frac{A}{\mu+1} t^{(\mu+1)(k-1)}x^{\mu+1} +\frac{1}{2}(k-1)^{2}x^{2}
\]
where the constant $A= \left( \frac{1-k}{\sqrt{c_{3}}} \right)^{\mu+3}$. This is the Case 4 in Table 1 of \cite{Muatje 2011}.

We note also that for $k=\frac{\mu+3}{\mu+1}$ the function $\omega(t)=A=const$. This recovers the second subcase of Case 1 in Table 1 of \cite{Muatje 2011} which is the Case 5.2 of \cite{Khalique 2008}.

\bigskip

We conclude that the seven cases 1-7 found in Table 1 of \cite{Muatje 2011} are just subcases of the above two general cases a) and b). To compare with these results one may adopt the notation $\omega=f$, $k=n$ and $\mu=p$.

\section{Conclusions}

\label{conclusions}

The purpose of the present work was to compute
the QFIs of time-dependent dynamical systems of the form $\ddot{q}^{a}= -\Gamma _{bc}^{a}\dot{q}^{b}\dot{q}^{c} -\omega(t) Q^{a}(q)$, where the connection coefficients are computed from the kinetic metric, using the direct method instead of the Noether symmetries as it is usually done. In the direct method one assumes that the QFI is of the form $I=K_{ab}\dot{q}^{a}\dot{q}^{b}+K_{a}\dot{q}^{a}+K$ and demands that $dI/dt=0$. This leads to a system of PDEs whose solution provides the QFIs. One key result is that the tensor $K_{ab}$ is a KT of the kinetic
metric.

We have discussed the solution of the system of equations at two levels. The first level is purely geometric and concerns the KT $K_{ab}$; and the second level is the physical one which concerns the quantities $\omega(t), Q^{a}(q)$  defining the dynamical system.

Concerning the first level we have applied two different methods:\newline
a. The polynomial method in
which one assumes a general polynomial form in the variable $t$ both for the
KT $K_{ab}$ and for the vector $K_{a}$. \newline
b. The basis method where one computes first a basis of the
KTs of order 2 of the kinetic metric and then expresses $K_{ab}$ in this basis assuming that the `components' are functions of $t$. \newline
In both methods the key point is to compute the scalar $K$.

Concerning the dynamical quantities $\omega(t), Q^{a}(q)$ we have chosen to
work in two ways:\newline
a. First we considered the polynomial method and assumed the
function $\omega(t)$ to be a polynomial leaving the quantities $Q^{a}$
unspecified. It is found that in this case the resulting dynamical system
admits two independent QFIs whose explicit expression together with
conditions involving the quantities $Q^{a}$ and the collineations of the
kinetic metric are given in Theorem \ref{thm.polynomial.omega}. \newline
b. In the basis method we worked the other way. That is, we assumed the quantities $Q^{a}(q)$ to be given by the time-dependent generalized Kepler potential $V=-\frac{\omega (t)}{r^{\nu }}$ and determined the functions $\omega (t)$ for which QFIs exist. The results of this detailed study are displayed in Table \ref{T2} for all values of $\nu$. For the values $\nu =-2, 1, 2$ we recovered the known results
concerning the time-dependent 3d oscillator, the time-dependent Kepler potential and the Newton-Cotes potential respectively. We note that these latter results have appeared over the years in many works whereas in the present discussion occur as particular cases of a single geometric approach.

The last part of our considerations concerns the well-known proposition that under a reparameterization the linear damping $\phi(t)\dot{q}^{a}$ can be absorbed to a time-dependent generalized force. We used this proposition in the case of a 1d non-linear second order time-dependent differential
equation, we determined the condition that the time-dependent coefficients of the equation must satisfy in order a QFI to exist and we computed this QFI. As an application we studied the properties of the well-known generalized Lane-Emden equation.

We note that one is possible to consider other dynamical quantities and/or
kinetic metric and compute the QFIs. What is the same in all cases is the
method of work which we hope we have presented adequately in the present
work.

\section{Appendix}

\label{sec.appendix}

Substituting the polynomial function $\omega(t)$ given by (\ref{pol}) in the system of equations (\ref{eq.red2a}) - (\ref{eq.red2e}) we have the following cases.
\bigskip

\underline{\textbf{I. Case $\mathbf{n=m}$}} (both $n$, $m$ finite) \bigskip

From equation (\ref{eq.red2a}) we obtain
\begin{equation}
C_{(k)ab} = -L_{(k-1)(a;b)}, \enskip k=1,...,n, \enskip L_{(n)(a;b)} =0. \label{eq.polc0}
\end{equation}
Therefore $L_{(n)a}$ is a KV of $\gamma_{ab}$.

Condition (\ref{eq.red2d}) gives
\begin{eqnarray*}
0 &=& -2\left( b_{1} +2b_{2}t + ... +\ell b_{\ell}t^{\ell-1} \right) \left(C_{(0)ab}Q^{b} +C_{(1)ab}Q^{b} t + ... + C_{(n)ab} Q^{b} \frac{t^{n}}{n}\right) + 2L_{(2)a}+6L_{(3)a}t +... + \\
&& + n(n-1)L_{(n)a}t^{n-2} -2\left( b_{0}+b_{1}t+ ... + b_{\ell}t^{\ell} \right) \left( C_{(1)ab}Q^{b} + C_{(2)ab}Q^{b}t +... + C_{(n)ab}Q^{b} t^{n-1} \right)+ \\
&& + \left( b_{0}+b_{1}t+ ... + b_{\ell}t^{\ell} \right) \left[ \left( L_{(0)b}Q^{b}\right) _{,a} + \left( L_{(1)b}Q^{b} \right)_{,a}t +... + \left( L_{(n-1)b}Q^{b} \right)_{,a}t^{n-1} + \left( L_{(n)b}Q^{b} \right)_{,a}t^{n} \right].
\end{eqnarray*}
This is a polynomial of the general form $P_{(0)a}(q) + P_{(1)a}(q)t +... +P_{(n+\ell)a}(q)t^{n+\ell}=0$. The vanishing of the coefficients $P_{(k)a}(q)$ in the last polynomial implies that
\begin{equation}
L_{(n)a}Q^{a}=s=const \label{eq.polc1}
\end{equation}
\begin{equation}
\sum_{s=0}^{\ell-1}\left[  -\frac{2(k+s)b_{(k+s\leq\ell)}}{n-s} C_{(n-s\geq0)ab}Q^{b} -2b_{(k+s\leq\ell)} C_{(n-s>0)ab}Q^{b} + b_{(k+s\leq\ell)} \left( L_{(n-s-1\geq0)b}Q^{b}\right)_{,a} \right]=0 \label{eq.polc2}
\end{equation}
where $k=1,2,...,\ell$,
\begin{equation}
-\sum_{s=1}^{\ell}\left[ \frac{2sb_{s}}{n-s} C_{(n-s\geq0)ab}Q^{b} \right] + \sum_{s=0}^{\ell} \left[ -2b_{s} C_{(n-s>0)ab}Q^{b} + b_{s} \left(L_{(n-s-1\geq0)b}Q^{b} \right)_{,a} \right]=0 \label{eq.polc3}
\end{equation}
and
\begin{equation}
k(k-1)L_{(k)a} - \sum_{s=1}^{\ell} \left[ \frac{2sb_{s}}{k-s-1} C_{(k-s-1\geq0)ab}Q^{b} \right] + \sum_{s=0}^{\ell} \left[ -2b_{s}C_{(k-s-1>0)ab}Q^{b} +b_{s}\left( L_{(k-s-2\geq0)b}Q^{b} \right)_{,a} \right] =0 \label{eq.polc4}
\end{equation}
where $k=2,3,...n$.

We note that in the $n+\ell+1$ formulae (\ref{eq.polc2}) - (\ref{eq.polc4}), when the undefined quantity $\frac{C_{(0)ab}}{0}$ appears in the calculations, it must be replaced by $C_{(0)ab}$ in order to have a consistent result.

We continue with the remaining constraints (\ref{eq.red2b}) and (\ref{eq.red2c}) in order to determine the scalar coefficient $K(t,q)$.

The solution of (\ref{eq.red2c}) is
\begin{eqnarray*}
K_{,t}&=& L_{(0)a}Q^{a} \left( b_{0} +b_{1}t +... +b_{\ell}t^{\ell} \right) + L_{(1)a}Q^{a} \left( b_{0}t +b_{1}t^{2} +... +b_{\ell}t^{\ell+1} \right) + ... + \\
&& + L_{(n-1)a}Q^{a} \left( b_{0}t^{n-1} +b_{1}t^{n} +... +b_{\ell}t^{n+\ell-1} \right) + s \left( b_{0}t^{n} +b_{1}t^{n+1} +... +b_{\ell}t^{n+\ell} \right) \implies \\
K &=& L_{(0)a}Q^{a} \left( b_{0}t +b_{1}\frac{t^{2}}{2} +... +b_{\ell}\frac{t^{\ell+1}}{\ell+1} \right) + L_{(1)a}Q^{a} \left( b_{0}\frac{t^{2}}{2} +b_{1}\frac{t^{3}}{3} +... +b_{\ell}\frac{t^{\ell+2}}{\ell+2} \right) + ... + \\
&& + L_{(n-1)a}Q^{a} \left( b_{0}\frac{t^{n}}{n} +b_{1}\frac{t^{n+1}}{n+1} +... +b_{\ell}\frac{t^{n+\ell}}{n+\ell} \right) + s \left( b_{0}\frac{t^{n+1}}{n+1} +b_{1}\frac{t^{n+2}}{n+2} +... +b_{\ell}\frac{t^{n+\ell+1}}{n+\ell+1} \right) +G(q).
\end{eqnarray*}
Replacing $K$ in (\ref{eq.red2b}) and using the conditions (\ref{eq.polc1}) - (\ref{eq.polc4}) we find that
\[
G_{,a}= 2b_{0}C_{(0)ab}Q^{b} -L_{(1)a}.
\]
Condition (\ref{eq.red2e}) is satisfied trivially from the above solutions.

The QFI is
\begin{eqnarray*}
I &=& \left( \frac{t^{n}}{n}C_{(n)ab} + ... + tC_{(1)ab} + C_{(0)ab} \right) \dot{q}^{a}\dot{q}^{b} + t^{n}L_{(n)a} \dot{q}^{a} + ... + tL_{(1)a}\dot{q}^{a} + L_{(0)a}\dot{q}^{a} + \\
&& + L_{(0)a}Q^{a} \left( b_{0}t +b_{1}\frac{t^{2}}{2} +... +b_{\ell}\frac{t^{\ell+1}}{\ell+1} \right) + L_{(1)a}Q^{a} \left( b_{0}\frac{t^{2}}{2} +b_{1}\frac{t^{3}}{3} +... +b_{\ell}\frac{t^{\ell+2}}{\ell+2} \right) + ... + \\
&& + L_{(n-1)a}Q^{a} \left( b_{0}\frac{t^{n}}{n} +b_{1}\frac{t^{n+1}}{n+1} +... +b_{\ell}\frac{t^{n+\ell}}{n+\ell} \right) + s \left( b_{0}\frac{t^{n+1}}{n+1} +b_{1}\frac{t^{n+2}}{n+2} +... +b_{\ell}\frac{t^{n+\ell+1}}{n+\ell+1} \right) +G(q)
\end{eqnarray*}
where $C_{(0)ab}$ is a KT, the KTs $C_{(k)ab} = -L_{(k-1)(a;b)}$ for $k=1,...,n$,  $L_{(n)a}$ is a KV such that $L_{(n)a}Q^{a}=s$, $G_{,a}= 2b_{0}C_{(0)ab}Q^{b} -L_{(1)a}$ and the conditions (\ref{eq.polc2}) - (\ref{eq.polc4}) are satisfied.
\bigskip

\underline{\textbf{II. Case $\mathbf{n \neq m}$}.} (one of $n$ or $m$ may be infinite)
\bigskip

We find QFIs that are subcases of those found in \textbf{Case I} and \textbf{Case III} which follows.

\bigskip

\underline{\textbf{III. Both $\mathbf{n}$, $\mathbf{m}$ are infinite.}}
\bigskip

In this case we consider the solution to have the form
\[
K_{ab}(t,q) = g(t)C_{ab}(q), \enskip K_{a}(t,q)= f(t)L_{a}(q)
\]
where the functions $g(t), f(t)$ are analytic so that
they may be represented by polynomial functions as follows
\begin{equation*}  \label{eq.thm1}
g(t) = \sum^n_{k=0} c_k t^k = c_0 + c_1 t + ... + c_n t^n
\end{equation*}
\begin{equation*}  \label{eq.thm2}
f(t) = \sum^m_{k=0} d_k t^k = d_0 + d_1 t + ... + d_m t^m.
\end{equation*}
In the above expressions the coefficients $c_{0}, c_{1}, ..., c_{n}$ and $d_{0}, d_{1}, ..., d_{m}$ are arbitrary constants. We find that only the following subcase gives a new independent FI. All other subcases give results already found.

\textbf{\underline{Subcase $\mathbf{(g = e^{\lambda t}}$, $\mathbf{f = e^{\mu t})}$, $\mathbf{\lambda \mu \neq 0}$}}.

In this case the system of equations (\ref{eq.red2a}) - (\ref{eq.red2d}) becomes (equation (\ref{eq.red2e}) is
satisfied trivially from the solutions found below):
\begin{eqnarray}
\lambda e^{\lambda t} C_{ab} + e^{\mu t} L_{(a;b)} &=& 0 \label{eq.polc5.1} \\
-2\left( b_{0}+b_{1}t + ... + b_{\ell}t^{\ell} \right) e^{\lambda t} C_{ab} Q^{b} + \mu e^{\mu t} L_a + K_{,a} &=& 0 \label{eq.polc5.2} \\
K_{,t} - (b_{0}+b_{1}t +... + b_{\ell}t^{\ell}) e^{\mu t} L_a Q^{a} &=& 0 \label{eq.polc5.3} \\
-2\left( b_{1} + 2b_{2}t + ... + \ell b_{\ell} t^{\ell-1} \right)e^{\lambda t} C_{ab}Q^{b} - 2 \lambda (b_{0}+b_{1}t+ ... + b_{\ell}t^{\ell})e^{\lambda t} C_{ab} Q^{b} + && \notag \\
+\mu^2 e^{\mu t} L_a + (b_{0}+b_{1}t + ... + b_{\ell} t^{\ell})e^{\mu t} \left( L_b Q^{b}\right)_{,a} &=& 0 \label{eq.polc5.4}.
\end{eqnarray}

We consider the following subcases.

a. \underline{For $\lambda \neq \mu$:}

From (\ref{eq.polc5.1}) we have that $C_{ab} =0$ and $L_{a}$ is a KV.

From (\ref{eq.polc5.4}) we find that $L_{a}=0$.

Therefore, the QFI $I_{e}(\lambda \neq \mu) =const$ which is trivial.
\bigskip

b. \underline{For $\lambda = \mu$:}

From (\ref{eq.polc5.1}) we have that $C_{ab} = - \frac{1}{\lambda} L_{(a;b)}$. Therefore $L_{(a;b)}$ is a KT.

We consider two cases according to the degree $\ell$ of the polynomial $\omega(t)$.
\bigskip

- Case $\ell=1$.

From (\ref{eq.polc5.4}) we find that
\begin{eqnarray}
\left( L_b Q^{b}\right)_{,a} &=& 2\lambda C_{ab}Q^{b} \label{eq.polc6.1} \\
\lambda^2 L_a + b_{0}\left( L_b Q^{b}\right)_{,a} - 2 (b_{1} + \lambda b_{0}) C_{ab} Q^{b} &=& 0. \label{eq.polc6.2}
\end{eqnarray}
Replacing with $C_{ab} = - \frac{1}{\lambda} L_{(a;b)}$ and by substituting (\ref{eq.polc6.1}) in (\ref{eq.polc6.2}) we obtain
\begin{eqnarray}
\left( L_b Q^{b}\right)_{,a}&=& -2 L_{(a;b)} Q^{b} \label{eq.polc6.3} \\
\lambda^3 L_a + 2b_{1} L_{(a;b)} Q^{b} &=& 0. \label{eq.polc6.4}
\end{eqnarray}

The solution of (\ref{eq.polc5.3}) is
\begin{equation*}
K=\left( \frac{b_{0}}{\lambda } - \frac{b_{1}}{\lambda^{2}} \right) e^{\lambda t}L_{a}Q^{a} + \frac{b_{1}}{\lambda} te^{\lambda t} L_{a}Q^{a} +G(q)
\end{equation*}%
which when replaced in (\ref{eq.polc5.2}) gives $G_{,a}=0$, that is $G=const\equiv0$.

The QFI is
\begin{equation}
I_{e}(\ell=1) = - e^{\lambda t} L_{(a;b)} \dot{q}^{a}\dot{q}^{b} + \lambda e^{\lambda t} L_{a}\dot{q}^{a} + \left( b_{0} - \frac{b_{1}}{\lambda} \right) e^{\lambda t}L_{a}Q^{a} + b_{1}te^{\lambda t} L_{a}Q^{a} \label{eq.polc6.4a}
\end{equation}
where $L_{(a;b)}$ is a KT, $\left( L_b Q^{b}\right)_{,a}= \frac{\lambda^{3}}{b_{1}}L_{a}$ and $\lambda^3 L_a = -2b_{1} L_{(a;b)} Q^{b}$.

- Case $\ell>1$.

From (\ref{eq.polc5.4}) we find that $\left( L_b Q^{b}\right)_{,a}= 2\lambda C_{ab}Q^{b}$, $C_{ab}Q^{b}= 0$ and $\lambda^2 L_a = 2b_{1}C_{ab} Q^{b}$.

Therefore $L_{a}=0$ and hence $C_{ab} = - \frac{1}{\lambda} L_{(a;b)}=0$. We end up with a trivial FI $I_{e}=const$.

\end{document}